\documentclass[12pt]{article}

\usepackage{amsmath}
\usepackage{amsfonts}
\usepackage{amssymb}
\usepackage{amsthm}
\usepackage{setspace}
\usepackage{color}
\usepackage{subcaption}
\usepackage{caption}
\usepackage{enumerate}
\usepackage[hidelinks]{hyperref}
\usepackage{graphicx}% Include figure files
\usepackage[hidelinks]{hyperref} %[linktocpage=true]
\usepackage[nottoc]{tocbibind}
\usepackage[utf8]{inputenc} 

\newcommand\ee{\end{equation}}
\newcommand\be{\begin{equation}}
\newcommand\eea{\end{eqnarray}}
\newcommand\bea{\begin{eqnarray}}

\newcommand\mpl{M_{\rm pl}}
\newcommand\comment[1]{}
\newcommand\expect[1]{\left\langle #1 \right\rangle}
\newcommand\bsb{\boldsymbol}
\comment{
\hypersetup{
colorlinks=true,
citecolor=DarkBlue,
linkcolor=DarkBlue,
urlcolor=DarkBlue,
}}

\def\O{\mathcal{O}}
\def\Tr{{\rm Tr}}
\def\d{\partial}

\def\E{{\mathcal{E}}}

\def\H{{\mathcal{H}}}

\def\L{{\mathcal L}}
\def\vep{\varepsilon}
\def\zrms{\zeta_{\rm rms}}

\def\r{\bsb r}

\def\k{{\bsb k}}

\def\vphi{\varphi}

\def\ep{\epsilon}
\def\H{\mathcal{H}}

\begin{document}

\begin{center}

  {\Large\bf Shapes of non-Gaussianity in warm inflation}

\vskip 1 cm
{Mehrdad Mirbabayi$^{a}$ and Andrei Gruzinov$^{b}$ }
\vskip 0.5 cm

{\em $^a$ International Centre for Theoretical Physics, Trieste, Italy}

{\em $^b$ New York University, New York, NY, USA}
\vskip 1cm

\end{center}
\noindent {\bf Abstract:} {\small {\em Sphaleron heating} has been recently proposed as a mechanism to realize warm inflation when inflaton is an axion coupled to pure Yang-Mills. As a result of heating, there is a friction coefficient $\gamma \propto T^3$ in the equation of motion for the inflaton, and a thermal contribution to cosmological fluctuations.
Without the knowledge of the inflaton potential, non-Gaussianity is the most promising way of searching for the signatures of this model. Building on an earlier work by Bastero-Gil, Berera, Moss and Ramos, we compute the scalar three-point correlation function and point out some distinct features in the squeezed and folded limits. As a detection strategy, we show that the combination of the equilateral template and one new template has a large overlap with the shape of non-Gaussianity over the range $0.01 \leq \gamma/H\leq 1000$, and in this range $0.7 <|f_{\rm NL}| < 50$. }

\vskip 1 cm

\section{Introduction}
Inflation is usually taught as a mechanism that erases subhorizon structures and creates (quantum mechanically) superhorizon ones. However, the idea that direct coupling to the rolling inflaton could lead to repeated particle production well within the horizon dates back to the 80's and 90's \cite{Moss,Maeda,Fang}. And for sizable couplings, it seems natural to expect that these particles thermalize. As such, {\em warm inflation} is conceptually different from {\em cold inflation} by having a non-decaying radiation component, a friction term in the background equation of motion for the inflaton
\be\label{bg}
\ddot\phi_0+(3 H +\gamma) \dot\phi_0 + V'(\phi_0) = 0,
\ee
and super-horizon fluctuations that are sourced by thermal noise. Of course, it is also phenomenologically different, and has motivated several works such as \cite{Graham_09,Bastero-Gil_11,Matias,Bastero-Gil} on its predictions for the cosmological correlation functions.

What motivates us to redo this analysis is a recent proposal called ``minimal warm inflation'' that seems to be a flawless microscopic realization of the idea \cite{Berghaus}. In this model, inflaton $\phi$ is an axion field and the source of particle production is its coupling to the Yang-Mills theory, 
\be\label{DL}
\Delta \L = \frac{\alpha\phi}{16 \pi f} \Tr[G_{\mu\nu} \tilde G^{\mu\nu}],
\ee
where $\alpha= \frac{g_{\rm YM}^2}{4\pi}$ and $\tilde G^{\mu\nu}= \vep^{\mu\nu\alpha\beta}G_{\alpha\beta}$. At temperatures much larger than the critical temperature of the gauge theory ($T\gg T_c$), the YM coupling is weak. Therefore the instanton effects, which contribute to the axion potential, are suppressed exponentially in $1/\alpha$. On the other hand, heating is believed to be via classical transitions called {\em sphalerons}. They are suppressed only by powers of $\alpha$:
\be
\gamma \sim \alpha^5 \frac{T^3}{f^2}.
\ee
Hence, minimal warm inflation evades the general concern \cite{Yokoyama} that thermal contributions to the inflaton potential might be too large to allow any other effect to be visible. We will briefly review this model in section \ref{sec:sph}, mainly to emphasize some qualitative features of sphalerons that make this proposal different from its Abelian cousins.

From a theoretical perspective, it is quite motivated to consider inflaton to be an axion field because its approximate shift symmetry can justify the flatness conditions on $V(\phi)$ needed for inflation to last long enough. Axions are generically coupled to gauge fields via the interaction \eqref{DL}, and it is plausible that a large range of model parameters and initial conditions might be attracted to warm inflation \cite{DeRocco}. Then the resulting friction would also help inflation last longer. Nevertheless, the ultimate way to test this idea is through its predictions for the cosmological fluctuations.

At first sight, working out these predictions might seem like a daunting task. Our analytic understanding of many aspects of YM theory at finite temperature is qualitative, and quantitative numerical analyses are expensive even without including the coupling to the inflaton and curved geometry. However, the task turns out to be relatively straightforward and the results remarkably universal because of the hierarchy between the scale of sphaleron heating $T$ and the curvature scale $H$ at which cosmological perturbations freeze out. These perturbations can be studied in an effective long wavelength description that consists of the inflaton, radiation fluid and gravity. The success of warm inflation is to leave few imprints of its underlying mechanism in this effective description. 

In fact, this inflaton-radiation effective field theory (EFT) was developed in \cite{Bastero-Gil}, well before the advent of the minimal warm inflation. In this EFT a key role in generation of cosmic perturbations is played by a noise term whose variance is fixed by the fluctuation-dissipation theorem. In section \ref{sec:EFT}, we review this effective description and identify the origin of the noise in our case as the fluctuations of $\Tr[G\tilde G]$ due to sphalerons. In particular, this noise has non-Gaussian correlators and this will be a source of non-Gaussianity in the cosmological perturbations. 

We study the linear cosmological perturbations in section \ref{sec:lin}, working at leading order in the slow-roll parameters and $H/T$. Given $V(\phi)$, we can then compute the scalar tilt and with some bits of UV data the tensor-to-scalar ratio. As an example, we will present a warm $\phi^4$ inflation coupled to $SU(2)$ that is consistent with the current $n_s-r$ constraints. 

\begin{figure}[t]
\centering
\includegraphics[scale =1.]{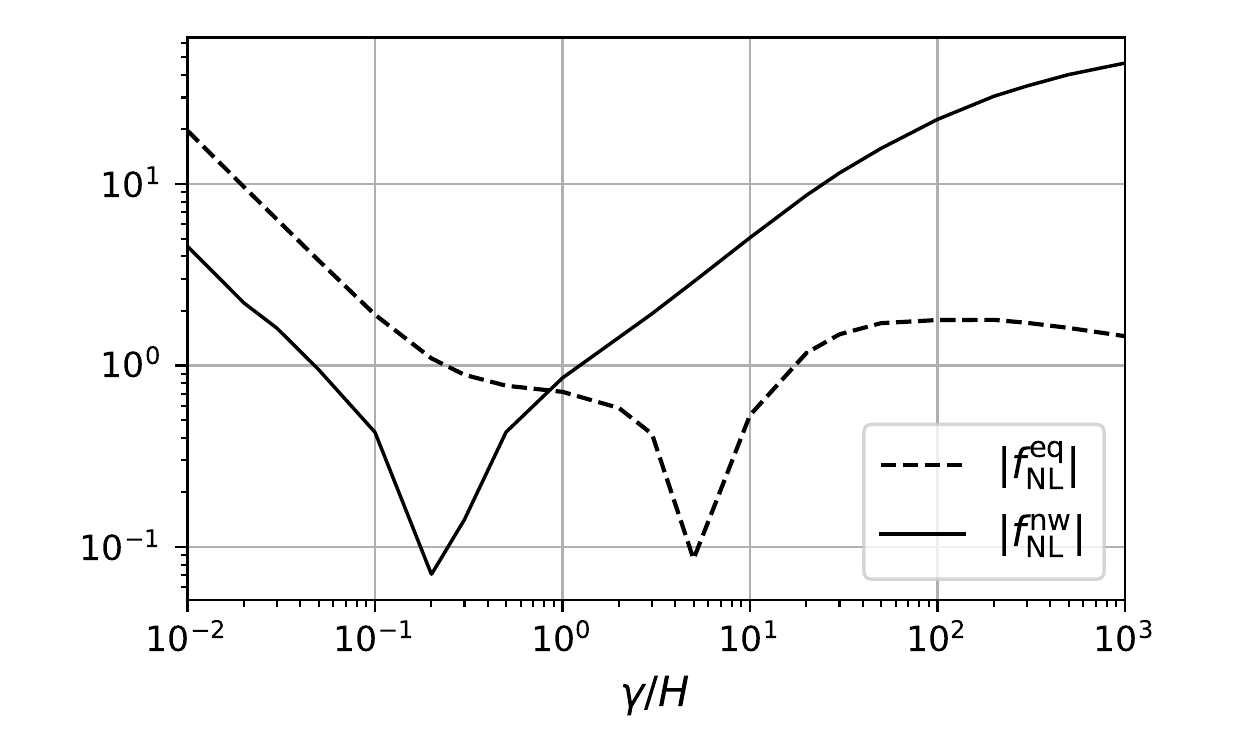} %[width= 11cm,height=8cm] 
\caption{\small{The coefficient for expanding the universal contribution to the bispectrum in terms of the equilateral and new-warm templates. They are both positive at large $\gamma/H$.}}
\label{fig:fs}
\end{figure}

Sadly, we do not know the real $V(\phi)$ and without this knowledge non-Gaussianity seems to be the only way (and a promising one) to find evidence for warm inflation. Indeed, there is a contribution originating from the nonlinear evolution of perturbations that is remarkably universal. It is fixed by $\gamma/H$ up to $\O(H/T)$ corrections. In section \ref{sec:est}, we will comment on some features of the non-Gaussianity, such as squeezed-limit behavior, UV sensitivity of the initial state, and parity violation. In section \ref{sec:shapes}, we work out the explicit predictions of the model for the 3-point correlator (bispectrum). The analysis is heavily numerical, but we will propose a new analytic template (new-warm) plus the well-known equilateral template as a basis with a considerable correlation with the numerical result for $0.01\leq \gamma/H \leq 1000$. As an example, figure \ref{fig:fs} shows the coefficients for expanding the universal contribution to the bispectrum (called $B_{211}$) in terms of these two shapes. 

We will conclude in section \ref{sec:con} with comments on the similarity and differences of this model with respect to the other particle production scenarios.

%%%%%%%%%%%%%%%%%%%%%%%%%%%%%%%%%%%%%%%%%%%%%%%%                                              
%%%%%%%%%%%%%%%%%%%%%%%%%%%%%%%%%%%%%%%%%%%%%%%           

%%%%%%%%%%%%%%%%%%%%%%%%%%%%%%%%%%%
\section{Sphaleron heating}\label{sec:sph}
There is a simple analogy that could be helpful in understanding how sphaleron heating works. Consider a pendulum with the Lagrangian
\be
L = \frac{1}{2} m R^2 \dot\theta^2 - m g R (1- \cos\theta),
\ee
and suppose we couple it to an external source $\vphi$,
\be
\Delta L =- \vphi \dot\theta.
\ee
For a constant $\vphi$, this term is of no consequence in classical physics though it matters quantum mechanically, much like the theta-term in QCD.\footnote{Our understanding of sphalerons (including this toy model) was mostly developed in conversations with Giovanni Villadoro. We thank Mark Trodden for pointing out earlier appearances of this analogy \cite{Arnold_strike,Trodden}.} When $\dot\vphi \neq 0$, we are exerting a torque on the pendulum, but for small $|\dot\vphi|$ this is not enough to turn it around a full cycle. Hence, the net flux of energy over a long time remains zero. 

Now suppose the pendulum is placed in a thermal bath of temperature $T$. At zero $\dot\vphi$, it would randomly, though rarely if $T \ll mgR$, complete full cycles $\theta\to \theta\pm 2\pi$. It follows that
\be
q(t) \equiv \int_0 ^t dt' \dot\theta(t')
\ee
performs a random-walk motion around the origin:
\be
\expect{q(t)^2} = 2 D t. 
\ee
If we now turn on a small $\dot\vphi \neq 0$, the jumps will be preferentially aligned with the torque. The fluctuation-dissipation theorem (FDT) implies that for $\dot\vphi\ll T$
\be
\expect{\dot\theta} \approx D \frac{\dot\vphi}{T},
\ee
and hence there is a net flux of energy into the heat bath, given by
\be
\dot E \approx \frac{D}{T} \dot\vphi^2.
\ee
This example shows the key role played in the heating mechanism by the periodicity of $\theta$ and the thermal transitions over the barrier that complete this cycle. These transitions are often called sphaleron transitions, best known for their potential role in baryogenesis \cite{Rubakov_sph}. There is an analog of the variable $q$ in four-dimensional Yang-Mills theory, placed in a finite box of volume $V$:
\be\label{Q}
Q(t) =\frac{\alpha}{16\pi} \int_0^t dt' \int_V d^3\r \Tr[G_{\mu\nu} \tilde G^{\mu\nu}]
\ee
As in our pendulum example, $Q$ is constructed using the time-derivative of a periodically identified quantity,
\be\label{Theta}
\frac{\alpha}{4\pi}\int_V d^3 \r \ \vep^{ijk} \Tr\left[A_i \d_j A_k -\frac{2}{3}ig_{\rm YM}A_i A_j A_k\right],
\ee
which can change by an integer under topologically nontrivial gauge transformations. At finite temperature, there are thermal fluctuations that complete the cycle and shift $Q$ by $\pm 1$. At weak coupling, these fluctuations (sphalerons) are well separated and hence approximately uncorrelated. Therefore, $Q(t)$ performs a random walk. The corresponding diffusion coefficient is related to the following limit of the thermal correlator
\be\label{Gammadef}
\lim_{k^0\to 0}\lim_{k^i\to 0}\frac{\alpha^2}{(16\pi)^2}\expect{\Tr[G_{\mu\nu} \tilde G^{\mu\nu}]_k \Tr[G_{\mu\nu} \tilde G^{\mu\nu}]_{k'}}= \Gamma (2\pi)^4\delta^4(k+k'),
\ee
which is equivalent to
\be
\lim_{t\to \infty}\lim_{V\to \infty}\expect{Q(t)^2} = \Gamma V t.
\ee
$\Gamma$ is called the sphaleron rate.

Much as in the example of the pendulum, in the presence of the coupling \eqref{DL}, a nonzero $\dot\phi$ induces an asymmetry in the jumps. When $T\gg \dot\phi/f$, by FDT \cite{Laine}
\be
\frac{\alpha}{16\pi} \expect{\Tr[G_{\mu\nu} \tilde G^{\mu\nu}]} \approx \frac{\Gamma}{2T} \frac{\dot\phi}{f},
\ee
which in turn implies a friction coefficient $\gamma = \Gamma/(2Tf^2)$ in the $\phi$ equation of motion \eqref{bg}, and a net flux of energy into the gluon bath. Note that it is important for the gauge group to be non-Abelian for the sphalerons to exist, and this mechanism would not work in the Abelian case. We will comment more on the latter in section \ref{sec:con}.

What is the temperature dependence of $\Gamma$? Let us expand \eqref{Gammadef} in the Hamiltonian basis to obtain
\be
\Gamma = \lim_{k^0\to 0}\lim_{k^i\to 0} \sum_{n,m} |\expect{n|O_k|m}|^2 e^{-E_n/T},\qquad  O \equiv \frac{\alpha}{16\pi }\Tr[G_{\mu\nu}\tilde G^{\mu\nu}].
\ee
Since YM theory has a mass gap $\sim T_c$, and $\expect{0|O_k|0} =0$ by translation invariance, the sphaleron rate is exponentially suppressed below the critical temperature. Well above it, by dimensional analysis, $\Gamma \propto T^4$. There remains a logarithmic dependence on $T_c/T$ through $\alpha\sim 1/\log(T/T_c)$. It has been argued in \cite{Arnold} that $\Gamma\propto \alpha^5 T^4$, because the characteristic size of sphalerons is $\sim 1/\alpha T$, but their time separation $\sim 1/\alpha^2 T$, and this expectation has been supported numerically \cite{Moore}. On the other hand, the corrections to the axion (inflaton) potential are suppressed by large powers of $T_c/T$. So it is enough to take $\alpha$ to be mildly less than one.

Our goal is to embed this setup in an inflating universe. Now there is a lower bound on $T$ from the horizon temperature $H$. In this paper, we will focus on the extreme case $T\gg H$ to be clearly distinct from the standard cold inflation. The fact that the characteristic scale of sphalerons is $\sim 1/T$ suggests that in this regime sphaleron heating operates almost the same as in the flat space, injecting energy at the rate $\gamma \dot\phi_0^2$ per unit volume into the gluon plasma. Therefore, the plasma density $\rho_0$, which at $T\gg T_c$ is to a good approximation that of a radiation gas, satisfies
\be\label{rhodot}
\dot\rho_0 + 4 H \rho_0 = \gamma \dot\phi_0^2. 
\ee
It is perhaps worth remarking that asking for $T\gg H$ by no means contradicts having $\rho_0\sim T^4\ll V$ and hence having inflation. Indeed, one can check that with $\gamma\propto T^3$, there is an {\em attractor} slow-roll solution 
\be\label{sr}
H^2\approx \frac{8\pi G}{3} V(\phi_0),\qquad \dot\phi_0 \approx -\frac{V'(\phi_0)}{3H+\gamma},\qquad 
\rho_0\approx \frac{\gamma}{4 H} {\dot\phi_0^2},
\ee
provided that the potential satisfies the following slow-roll conditions \cite{Moss_attractor}
\be
\frac{{V'}^2}{16\pi G V^2}\ll 1+\frac{\gamma}{3H} ,\qquad \frac{|V''|}{8\pi G V } \ll 1+\frac{\gamma}{3H}.
\ee
On this solution, background quantities such as $H$, $\gamma$, etc. change slowly and as a result the spectrum of cosmological perturbations will be approximately scale invariant.
%%%%%%%%%%%%%%%%%%%%%%%%%%%%%%%%%%%%%%%%%%
\section{Effective hydro description}\label{sec:EFT}
Warm inflation offers an alternative origin for the observed cosmological perturbations. These are perturbations that stretch to wavelengths $\lambda\gg 1/H$ during inflation. Hot gluons, having a mean-free-path $\ell \sim 1/\alpha T$, scatter several times over these scales. Hence, to study the horizon-crossing and superhorizon evolution of perturbations, we can switch to a hydrodynamic description. Since there are no internal conserved charges, this is just a radiation fluid with local energy density $\rho$, four velocity $u^\mu$, and energy-momentum tensor
\be
T^{\rm rad}_{\mu\nu} = \frac{4}{3}\rho u_\mu u_\nu +\frac{1}{3} \rho g_{\mu\nu}+ \cdots
\ee
Dots represent the corrections to the equation of state, suppressed by $\alpha_{\rm YM}$, or the higher derivative terms, suppressed by $\ell/\lambda$. We are going to neglect them for the most part. This eliminates all dissipative terms except for the sphaleron heating. The latter couples the radiation fluid to the inflaton and is essential to sustain warm inflation. Recalling that the friction term $\gamma \dot\phi$ was the linear response of $O = \frac{\alpha}{16\pi f}\Tr[G_{\mu\nu}\tilde G^{\mu\nu}]$ to a uniform $\dot\phi$, the covariant version of the equation of motion of $\phi$ at lowest order in this derivative expansion reads
\be\label{phieq}
D^2 \phi - V'(\phi) = \gamma u^\mu \d_\mu \phi + \xi,
\ee
where $D^2$ is the covariant d'Alembertian, $\gamma$ is a function of the local temperature (or, equivalently, of $\rho$):
\be
\gamma = \gamma_0 \left(\frac{\rho}{\rho_0}\right)^{3/4},
\ee
and $\xi$ is a noise whose presence is dictated by the FDT. From a microscopic point of view
\be
\xi = \frac{1}{f}(O -\expect{O}),
\ee
where $\expect{.}$ is the thermal average. At a qualitative level, we know that $\xi$ has a correlation length of order $\ell$. It then follows from \eqref{Gammadef} that, in the long-wavelength description, we can approximate 
\be\label{xi2pf}
\expect{\xi(x_1)\xi(x_2)} \approx 2\gamma T \delta^4(x_2 - x_1),
\ee
where we used 
\be\label{FDT}
\Gamma =  2 \gamma T f^2.
\ee
The same qualitative knowledge implies that higher-point connected correlators of $\xi$ can also be approximated as vanishing except at the coincident point.

At this point, the attentive reader might be skeptical (and rightly so) of the adequacy of this effective description for a reliable calculation of inflationary predictions. After all, every comoving mode we observe today would in this model have had a physical wavelength $\lambda\sim \ell$ at some early enough time, and could have been influenced by all sorts of thermal fluctuations as well as the sphalerons. Our long-wavelength approximation to the noise correlator \eqref{xi2pf} carries no information about those. 

In fact, for now we should think of this effective description merely as a framework to evolve a given initial state of perturbations that emerges from the microscopic scales $\sim \ell$. It requires a more quantitative analysis, presented in the next two sections, to see that this framework really is predictive despite what might be happening at those microscopic scales. 

Now, back to our long-wavelength model for the noise, on which hinges the above conclusion. Let us consider the weak coupling limit in which sphalerons are rare and identifiable. Suppose we smooth $O(x)$ over spacetime regions of 4-volume $v$ bigger than the sphaleron size, $v\gg \ell^4$, but smaller than their separation. In region $i$ the product $O_i v$ is small (thermal fluctuations average out) unless there is a sphaleron. For almost half of these $O_i v \approx +1$, and for the rest $O_i v \approx -1$. If we denote their spacetime densities respectively by $n_+$ and $n_-$, then the probability of finding one in a cell is $p_\pm = n_\pm v \ll 1$, and to leading order in $p_\pm$:
\be\label{discrete}
\expect{O_{i_1} O_{i_2}\cdots O_{i_N}} \approx [{n_+}+ (-)^N {n_-}]
\frac{\delta_{i_1i_2}\cdots\delta_{i_1 i_N}}{v^{N-1}}.
\ee
We can now identify (in the rest frame of gluon plasma and for small $\dot\phi$), $n_+ - n_- = \gamma f \dot\phi$, and $n_+ + n_- = \Gamma$. 

Macroscopically, \eqref{discrete} can be replaced with a continuous version, which after removing the mean and dividing by $f$ gives the following Poisson statistics for $\xi$
\be\label{xiN}
\expect{\xi(x_1)\cdots \xi(x_N)} \approx \frac{\Gamma}{f^N}\left(\prod_{i=2}^N \delta^4(x_i - x_1)\right)\times \left\{\begin{array}{cc}1, & \qquad \text{$N$ even,}\\[10pt]\frac{u^\mu\d_\mu\phi}{2f T},&\qquad \text{$N$ odd.}\end{array}\right.
\ee
This noise model is often denoted as the shot-noise. Its precise form is specific to the above microscopic picture. For instance, if there were two types of random process one shifting $O_i v$ by 1 unit and the other by 2 units, correlators of more than two $\xi$ would be different (larger). More realistically, one might expect \eqref{xiN} to become inaccurate when $\alpha \to 1$ because the gap between the sphaleron size and separation closes. However, in this limit $\Gamma \sim T^4$ and \eqref{xiN} would still be a reasonable estimate for the non-Gaussian noise correlators. 

Also our treatment of $\xi$ as a classical random field is justified quite generally because by FDT the commutator of two $O$, which is proportional to their retarded Green's function, is suppressed by $\omega/T$ compared to their variance.

Having specified the long-distance effective equation of motion for $\phi$, we can next derive that of the radiation fluid. Given that the coupling $\phi\Tr[G\tilde G]$ does not depend on the metric, the conservation of total energy and momentum implies
\be\label{fleq}
D^\mu T^{\rm rad}_{\mu\nu} =- \d_\nu \phi (\gamma u^\mu \d_\mu \phi + \xi).
\ee
At leading order in slow-roll parameters, equations \eqref{phieq} and \eqref{fleq} are all we need to evolve cosmological scalar perturbations. We will work at this order, which allows us to neglect the variation of $\dot\phi_0,\rho_0$ and $H$ when solving for the perturbations, and approximate the background metric by
\be
ds^2 = -dt^2 +a(t)^2 dx^2,\qquad a(t) = e^{Ht}.
\ee
In this approximation, we can also ignore scalar fluctuations of the metric, and, as will be further discussed, use the following simple relation for the gauge-invariant scalar perturbations \cite{Maldacena}
\be\label{zeta}
\lim_{k\to 0} \zeta_\k(t) = - \frac{H }{\dot\phi_0} \delta\phi_\k(t).
\ee
For the cosmological scales tensor perturbations can be treated in the standard way. They are hardly affected by the microscopic thermal fluctuations \cite{GW}. It is conceivable that the scalar fluctuations become sizable toward the end of inflation and at second order source the tensor modes. It would be interesting to investigate the stochastic gravitational wave signal of this model. See \cite{Qiu,Klose} for some related works. 

%%%%%%%%%%%%%%%%%%%%%%%%%%%%%%%%%%%%%%%%%%%%%%%%%%%%%%%%%%%%%%%%%%
\section{Linear perturbations}\label{sec:lin}
In this section, we will study the scalar perturbations at the linear level. With some hindsight, in order to simplify the non-Gaussianity analysis, we define the perturbations $P = (\E,\Psi,\Phi)$ via
\be\label{def}
\phi = \phi_0 + \dot\phi_0 \Phi,\quad
T^{~0}_{{\rm rad}~0} = -\rho_0 (1+ \E),\quad
\d_i T^{~i}_{{\rm rad}~0} =  - \frac{4\rho_0}{3a^2} \nabla^2 \Psi.
\ee
Substituting these in \eqref{phieq} and \eqref{fleq}, neglecting slow-roll suppressed terms proportional to $\ddot\phi_0$ and $V''(\phi_0)$, and using the background equations \eqref{sr} we find (in momentum space, and in terms of the conformal time $\eta = -\frac{1}{a H }$)
\be\label{linear}\begin{split}
  \E_\k ' -  \frac{\E_\k}{\eta} + \frac{4 H k^2\eta}{3} \Psi_\k - 8 H \Phi_\k' & = -\frac{4 }{\gamma\dot\phi\eta}  \xi_{\k}\\[10pt]
\Psi_\k'- 3 \frac{\Psi_\k}{\eta}- \frac{1}{4H\eta } \E_\k - \frac{3}{\eta} \Phi_\k &=0\\[10pt]
\Phi_\k'' - (2 + \gamma/H) \frac{\Phi_\k'}{\eta}+ \frac{3}{4H^2\eta^2} \gamma \E_\k + k^2 \Phi_\k
& =  - \frac{1}{\dot\phi H^2\eta^2} \xi_{\k},
\end{split}\ee
where prime denotes $d/d\eta$, and we have replaced $\gamma_0\to \gamma$, $\dot\phi_0\to \dot\phi$, which are understood to be the background values. The noise is an approximately classical variable with 2-point function in momentum space (see \eqref{xi2pf} and \eqref{FDT})
\be\label{xi22}
\expect{\xi_\k(\eta)\xi_{\k'}(\eta')} =  2\gamma T H^4\eta^4 \delta(\eta-\eta')(2\pi)^3\delta^3(\k + \k').
\ee
If $\Phi, \E$ and $\Psi$ were in a thermal state at the time of horizon-crossing, they would have a large occupation number and would be approximately classical. 

However, on the inflating background cosmological perturbations stretch from microscopic scales. To obtain the state of the perturbations at superhorizon scales, we should in principle start from the adiabatic vacuum at $\lambda\ll T$ and follow their evolution through the thermal and cosmological scales. The result certainly depends on the microscopic details. Our effective hydro description is predictive if the excitations at $\lambda \gg \ell\sim 1/T$ overwhelm the earlier ones. This is indeed the case, as we will now argue.

%%%%%%%%%%%%%%%%%%%%%%%%%%%%%
\subsection{Asymptotic behaviors} \label{sec:asymp}
We address this question with the help of the effective theory. Even though it breaks down at short scales, and the noise correlator is really a smeared delta function of width $\ell$, we can still rely on it for an order of magnitude estimate of the amount of emission at those scales. Let us define the retarded Green's functions $G(\eta,\eta')=(G_\E(\eta,\eta'),G_\Psi(\eta,\eta'),G_\Phi(\eta,\eta'))$ as the functions that vanish for $\eta<\eta'$ and satisfy \eqref{linear} with $\xi(\eta)\to \dot\phi \delta(\eta-\eta')$. Assuming the effective theory is valid at all scales, we can use \eqref{zeta} and \eqref{xi22} to write the sourced contribution to the scalar power as
\be\label{zxi}
k^3\expect{\zeta_{\k_1} \zeta_{\k_2}}'_\xi = \frac{H^2}{\dot\phi_0^2} 2\gamma T F_2(\gamma/H),
\ee
where prime on the correlation functions means that we dropped $(2\pi)^3$ times the momentum-conservation delta function, and 
\be\label{F2}
F_2(\gamma/H) = k^3 \int_{-\infty}^{0} d\eta \eta^4 G_{\Phi_\k}(0,\eta)^2.
\ee
Of course, the effective theory breaks down when $|k\eta|>1/H\ell$. However, we will see that this integral converges at scales determined by $H$ and $\gamma$. This a posteriori justifies our inaccurate description of the microscopic scales and extending the $\eta$ integral to $-\infty$. 

We can only solve for $G$ numerically, but the above convergence property can be inferred from its asymptotic behavior, which can be determined analytically. In the process, we will also justify the relation \eqref{zeta}.

So we begin by finding the superhorizon behavior of the four independent homogeneous solutions of \eqref{linear}. They have the following power-law behaviors up to $\O(k^2\eta^2)$ corrections
\be\label{super}\begin{split}
P_c &= \frac{1}{H}(0,-1,1),\\[10pt]
P_\pm & = \left(\frac{8\alpha_\pm}{\alpha_\pm -1},\frac{5\alpha_\pm -3}{H(\alpha_\pm-1)(\alpha_\pm -3)},\frac{1}{H}\right) (- H\eta)^{\alpha_\pm},\\[10 pt]
P_u & = (0,1,0)H^2\eta^3,
\end{split}\ee
where
\be
\alpha_\pm = \frac{1}{2 H}\left(4H+\gamma \pm \sqrt{4H^2 - 20\gamma H +\gamma^2}\right).
\ee
Since ${\rm Re} \alpha_\pm >0$, there is only one non-decaying mode, which corresponds to the perturbations of the clock along the attractor trajectory. $\zeta = - H \delta\phi/\dot\phi$ measures the extra expansion due to this time-shift.

The Green's function is a combination of the independent solutions. In the limit $k|\eta'|\ll 1$, we find 
\be
G(\eta,\eta') = \theta(\eta-\eta') \left[\frac{A_c}{H\eta'} P_c(\eta) + \frac{A_+}{{(-H\eta')}^{\alpha_+ +1}} P_+(\eta)
  +\frac{A_-}{{(-H\eta')}^{\alpha_- +1}} P_-(\eta)\right]
\ee
where
\be
A_c = \frac{4H}{3H+7\gamma},\qquad A_\pm = H\frac{\mp 5(H-\gamma) +2 \sqrt{4H^2 - 20\gamma H +\gamma^2}}{(3H+ 7\gamma)\sqrt{4H^2 - 20\gamma H +\gamma^2}}.
\ee
Given that only $P_c(0)\neq 0$, the late time behavior of \eqref{F2} is $\int^0 d\eta \eta^2$, which is finite.

In the subhorizon regime, when $k|\eta|\gg 1$, we can use the WKB approximation to find two positive frequency solutions:
\be\begin{split}
P_{\rm inf}(\eta) &= \left(18,\frac{15 i }{2 H k\eta} , \frac{1}{H}\right) (-k\eta)^{1+ \frac{\gamma}{2H}} e^{-i k\eta},\\[10pt]
P_{\rm sound}(\eta) &= \left(\frac{H}{\gamma},\frac{i\sqrt{3}}{4 \gamma k\eta} , -\frac{9}{8 H k^2 \eta^2}\right)
(-k\eta)^{5/2} e^{-\frac{i}{\sqrt{3}} k\eta},\end{split}
\ee
and two negative frequency ones by complex conjugation. In this limit, we find
\be%
G(\eta,\eta') =\theta(\eta-\eta')\left[ -\frac{i k e^{i k \eta'}}{2   (-k\eta')^{3+\frac{\gamma}{2H}}}P_{\rm inf}(\eta)
+ \frac{2 k e^{\frac{i}{\sqrt{3}}k\eta'}}{ (-k\eta')^{7/2}} P_{\rm sound}(\eta)+ {\rm c.c.}\right].
\ee
Generically, both $P_{\rm inf}(\eta)$ and $P_{\rm sound}(\eta)$ have a nonzero (though $\gamma$-dependent) overlap with $P_c(\eta)$. Therefore, the largest early-time contribution to \eqref{F2} goes as
\be\label{F2early}
F_2(\gamma/H)_{\rm early}\propto \int_{-\infty} d\eta (-\eta)^{-2-\gamma/H}
\ee
for $\gamma <H$ and as
\be
F_2(\gamma/H)_{\rm early}\propto \int_{-\infty} d\eta (-\eta)^{-3}
\ee
for $\gamma >H$, which are both finite. This justifies neglecting what happens when $\lambda \sim \ell$. In principle, the noise spectrum must be modified and the hydrodynamic picture discarded when $|\k\eta|\sim 1/H\ell$. However, taking it as a proxy for the true answer, we see that the contribution from those scales is suppressed by powers of $H\ell\ll 1$. 

Before moving to the implications of this result, it is useful for our future discussion of non-Gaussianity to analyze one more limit, namely $\gamma \gg H$. Now another scale appears in the evolution of linear perturbations and the above WKB solutions are no longer valid when $1\ll k|\eta|\ll \gamma/H$. The sound mode changes to
\be
\tilde P_{\rm sound}(\eta) = \left(1,\frac{i\sqrt{3}}{4H k\eta} , \frac{3 \sqrt{3} i}{4 H k \eta}\right)
(-k\eta)^{10} e^{-\frac{i}{\sqrt{3}} k\eta}.
\ee
On the other hand, since $\Phi$ is over-damped, there will be a fast decaying mode
\be
\Phi'' -\gamma \frac{\Phi'}{H\eta} \approx 0,
\ee
and a slow-decaying one with $\E = -12 H \Phi$ and 
\be\label{slow}
\gamma \eta \Phi_\k' + 9 \gamma \Phi_\k - H k^2 \eta^2 \Phi_\k \approx 0.
\ee
Hence, $P_{\rm inf}$ is replaced by 
\be\label{fastslow}\begin{split}
P_{\rm fast}(\eta) &=\left(8,\frac{5 }{\gamma},\frac{1}{H}\right)(-k\eta)^{\gamma/H},\\[10pt]
P_{\rm slow}(\eta) &=\left(-12,\frac{15 }{\gamma}-\frac{144}{Hk^2\eta^2},\frac{1}{H}\right)
(-k\eta)^{-9}\exp\left(\frac{H k^2 \eta^2}{2\gamma}\right).
\end{split}\ee
We see that $P_{\rm slow}$ rapidly decays prior to $k|\eta|\sim \sqrt{\gamma/H}$ and grows afterwards while the magnitude of all other modes decay with a large power of $\eta$. See \cite{Graham_09,Bastero-Gil_11} for earlier works that identify this growing mode. 

It follows that when $\gamma\gg H$ the main contribution to the $\zeta$ correlators comes from the excitation of the slow mode during the period $k|\eta| \sim \sqrt{\gamma/H}$, because they will grow by a factor of about $(\gamma/H)^{9/2}$ until horizon-crossing. In practice, the numerical solution does not match this analytic estimate unless $\gamma > 10^4 H$. However, it does demonstrate the same qualitative behavior for any large $\gamma/H$. We will see that the separation between the scale of excitation of perturbations and the scale at which they freeze leads to interesting features in the squeezed-limit three (and higher) point correlators.
%%%%%%%%%%%%%%%%%%%%%%%%%%%
\subsection{Scalar power, dominated by the stochastic emission}
\begin{figure}[t]
\centering
\includegraphics[scale = 1]{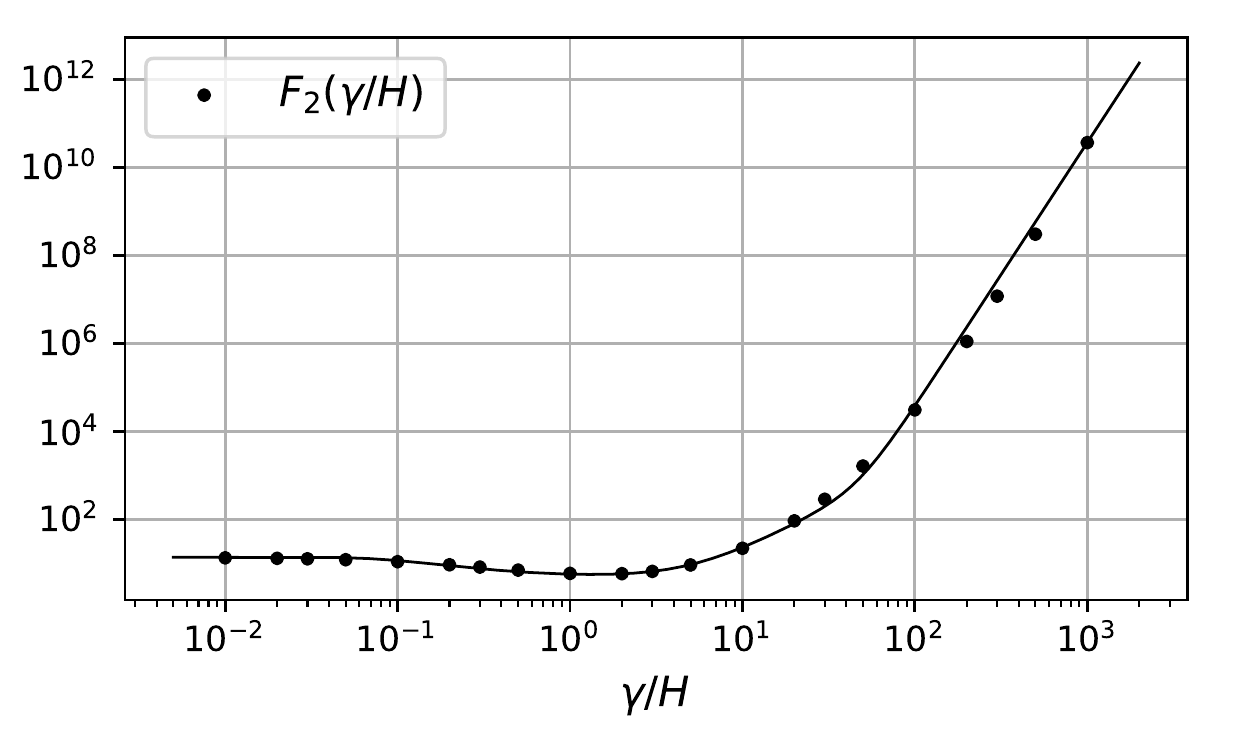} %[width= 11cm,height=8cm]                               
\caption{\small{Plot of $F_2(\gamma/H)$ and the fitting formula \eqref{F2fit}. }}
\label{fig:F2}
\end{figure}

We established within the hydro description that the sourced contribution to the scalar power spectrum \eqref{zxi} is peaked when the effective theory is valid, and the uncertainty from the short scales is suppressed by a power of $H/T$. A plot of $F_2(\gamma/H)$ obtained by numerical integration is given in figure \ref{fig:F2}, where it is compared with a fitting formula\footnote{This result and therefore our prediction for the scalar power noticeably deviates from eq. (79) of \cite{Graham_09}, eq. (4.27) of \cite{Bastero-Gil_11}, and eq. (3.29) of \cite{Bastero-Gil}.}
\be\label{F2fit}
F_2(x)\approx 3.7\times 10^{-8} x^6+14 \left[\frac{1}{3} \left(1+\frac{x^2}{25}\right)
  + \frac{2}{3} \tanh\left(\frac{1}{10 x}\right)\right],\qquad 10^{-2} \leq x \leq 10^3.
\ee
%%%%%%%%%%%%%%%%%%%%%%%%%%%%%%%%%%%%%%%%%%%%%%%%                                              
%%%%%%%%%%%%%%%%%%%%%%%%%%%%%%%%%%%%%%%%%%%%%%%  
It reaches its minimum at $x \approx 1.5 $ with $F_2(1.5) \approx 5.8$, and in the limiting case $x\to 0$
\be\label{F2lim}
F_2(0) \approx 13.9.
\ee
Therefore, we can approximate the full 2-point correlator of $\zeta$ by \eqref{zxi} except when $\gamma T < H^2$; for small enough $\gamma$, inflaton decouples and we should recover the single-field result. So we can write 
\be\label{zxi+vac}
P(k) = \expect{\zeta_{\k} \zeta_{-\k}}' \approx 
\frac{H^2}{2\dot\phi^2 k^3} [4\gamma T F_2(\gamma/H)+H^2].
\ee
This is clearly a good approximation if the sourced contribution is much larger than the standard vacuum fluctuations (the second term in the parenthesis). Otherwise, as long as we are working with linear perturbations, we can simply add the emission by the classical source to the quantum zero-point fluctuations.

Nevertheless, our main interest in this work is when $\zeta$ is dominated by the source, i.e.
\be\label{Qmin}
\frac{\gamma}{H} > \frac{H}{4 T F_2(0)}.
\ee
This might suggest that we can have arbitrarily small $\gamma/H$ by increasing $T/H$. However, in any given model, decreasing $\gamma$ leads to decreasing $T$ since $\gamma$ controls the rate of heating. This results in a minimum $\gamma/H$ for which $\zeta$ is noise-dominated. For instance, in minimal warm inflation with $SU(N)$ gauge group and $T\gg T_c$, we have
\be
\rho = \frac{\pi^2 g_*}{30} T^4,\qquad g_*=2(N^2 -1).
\ee
Using this in the quasi-equilibrium condition $4 H \rho = \gamma \dot\phi^2$ allows us to eliminate $\dot\phi$ and obtain 
\be\label{T(Y)}
\frac{T}{H} = \left(\frac{15 \gamma^2 F_2(\gamma/H)}{\pi^2 g_* H^2 \zeta_{\rm rms}^2}\right)^{1/3},
\ee
where we assumed the sourced contribution to \eqref{zxi+vac} dominates and matches the CMB observation \cite{Planck}
\be
\zeta_{\rm rms}^2\equiv k^3 \expect{\zeta_{\k_1} \zeta_{\k_2}}' \approx  4.1 \times 10^{-8}.
\ee
Comparing \eqref{T(Y)} with the condition \eqref{Qmin} for $\zeta_\xi > \zeta_{\rm vac}$ gives
\be\label{gamma_min}
{\rm min}\left(\frac{\gamma}{H}\right) \approx \left(\frac{\pi^2 g_* \zeta_{\rm rms}^2}{960 F_2^4(0)}\right)^{1/5}\qquad
(= 2.3\times10^{-3} \quad \text{for $SU(2)$}).
\ee
So there is a finite but considerable range of parameters in which $\gamma<H$. In this range warm inflation does not significantly modify the background evolution, but it does change the origin of perturbations.

It is an interesting curiosity to ask if the inflaton perturbations ever come into thermal equilibrium with the gluon plasma. Note that we never needed the answer to this question to determine the scalar power. Still, in the low friction regime, we can ask if the resulting power is consistent with a thermal spectrum at the horizon scale. The answer is no: from \eqref{zxi+vac} we can infer $\delta\phi_H \sim \sqrt{\gamma T}$, while thermal spectrum would imply $\delta\phi_H\sim \sqrt{HT}\gg \sqrt{\gamma T}$.

Could the fluctuations be thermalized at earlier times when their wavelength is $\sim 1/T$? Because the thermal occupation number at scales of order temperature is $\O(1)$, this would imply a UV contribution to the scalar power of order of a few times the vacuum contribution. The answer is again no. As argued above, the contribution to the scalar power from those scales can be estimated by the UV part of \eqref{F2}. As can be seen from \eqref{F2early}, in the $\gamma\ll H$ regime this is suppressed by $H/T$, giving $\zeta_{\xi,{\rm early}}/\zeta_{\rm vac} \sim \sqrt{\gamma/ H}\ll 1$. It is conceivable that an approximate thermal state is reached at early times and when $\gamma\sim H$ or larger. But those thermal fluctuations would be buried under many more that follow. 
%%%%%%%%%%%%%%%%%%%%%%%%%%%%%%%%%%%%%%%%%%%%
\subsection{An example: warm $\phi^4$ inflation}
\begin{figure}[t]
\centering
\includegraphics[scale = 1]{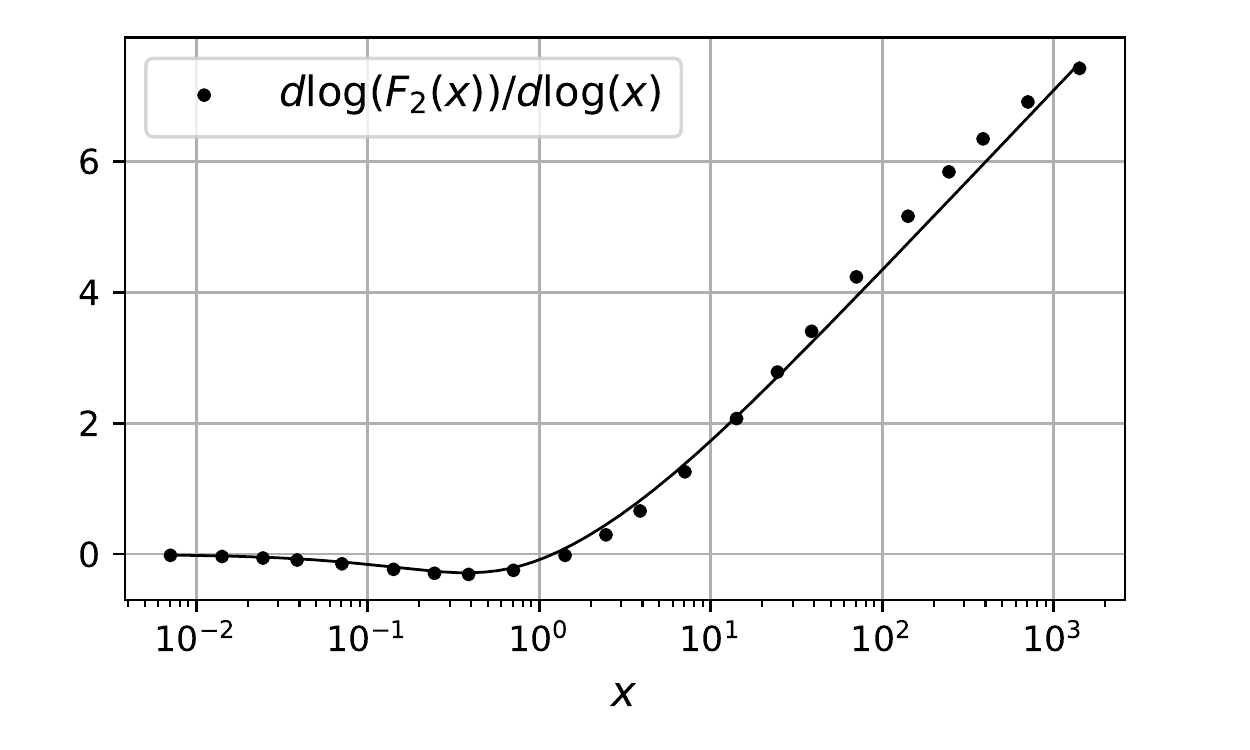} %[width= 11cm,height=8cm]                               
\caption{\small{Plot of $\frac{d\log(F_2(x))}{d\log(x)}$ and the fitting formula \eqref{dF2fit}. }}
\label{fig:F2_tilt}
\end{figure}

Many models of single-field inflation have been ruled out by their predictions for the scalar tilt and the tensor-to-scalar ratio. This includes all convex power-law potentials \cite{Planck}. Warm inflation alters these constraints by changing the scalar power. $m^2\phi^2$ model continues to be excluded, now because of wrong prediction for the tilt, but there is an acceptable range for $V(\phi) = \frac{\lambda}{4}\phi^4$. If we define
\be
\hat \phi \equiv \sqrt{\lambda} \phi,\qquad \hat T \equiv \gamma^{1/3},
\ee
and work in units where $\mpl^2 \equiv \frac{1}{8\pi G} = \frac{1}{\lambda}$, the background equations of motion read
\be\label{bgr}\begin{split}
\d_t^2\hat \phi &= - (3 H + \hat T^3)\d_t\hat\phi -\hat\phi^3,\\[10pt]
\d_t \hat T &= - H \hat T +\beta (\d_t\hat\phi)^2,\\[10pt]
H^2 &= \frac{1}{3} \left[\frac{1}{2} (\d_t \hat\phi)^2 + \frac{\hat\phi^4}{4}+ \frac{1}{4\beta} \hat T^4\right],
\end{split}\ee
where $\beta$ is a constant combination of model parameters (note: $\rho$ is radiation density)
\be\label{beta}
\beta = \frac{\gamma^{4/3}}{4 \lambda \rho}.
\ee
For large enough $\hat\phi$, the system \eqref{bgr} has an inflating attractor. At $\beta = 0$, radiation decouples and we are in the conventional cold inflation scenario; an initial excursion of $\hat\phi_i = \frac{\phi_i}{\mpl}\sim 22$ gives about 60 e-folds of slow-roll inflation. When $\beta\neq 0$, a smaller range of $\hat\phi$ would suffice. For instance, at $\beta = 10$ it takes about 60 e-folds to arrive along the attractor solution from $\hat \phi_i \sim 12$ to the reheating surface where $\rho \sim V(\phi)$. Given $\hat\phi(t),\hat T(t), H(t)$, we can then calculate the scalar tilt from \eqref{zxi+vac}
\be
1-n_s =2 \frac{\d_t^2 \hat \phi}{H\d_t \hat\phi} - 2 \frac{\dot H}{H^2}-4 \frac{\d_t \hat T}{H \hat T}
+\frac{d\log F_2(\gamma/H)}{d \log(\gamma/H)} \left(\frac{\dot H}{H^2} - 3\frac{\d_t \hat T}{H \hat T}\right)
\ee
where the logarithmic derivative of $F_2$ is computed numerically.\footnote{Figure \ref{fig:F2_tilt} shows a plot and comparison with the following fitting formula for $0.01\leq \gamma/H\leq 1000$
\be\label{dF2fit}
\frac{d\log F_2(x)}{d\log x} \approx \frac{6}{5}\left[(\log(2/3)+1) e^{-4 x}-1+\log\left(\frac{3}{2}+x\right)\right].
\ee}
%%%%%%%%%%%%%%%%%%%%%%%%%%%%%%%%%%%%%%%%%%%%%%%%                                              
%%%%%%%%%%%%%%%%%%%%%%%%%%%%%%%%%%%%%%%%%%%%%%%  
In particular, we obtain at $N_e = 55$ 
\be
\phi \approx 11.6 \mpl, \qquad \gamma \approx 5.34 H,\qquad H = 39.0 \sqrt{\lambda}\mpl,\qquad 1- n_s \approx 0.0337.
\ee
The absolute value of the temperature and the tensor-to-scalar ratio can be determined only after fixing the underlying model, and hence $g_*$. For $SU(2)$ gauge group, we obtain from \eqref{T(Y)} and \eqref{beta}
\be
\frac{T}{H} \approx 1210,\qquad r = \frac{4H^2}{\mpl^2\zrms^2}\approx 4.70 \times 10^{-7}.
\ee
These predictions for $n_s$ and $r$ lie well within the current bounds.
%%%%%%%%%%%%%%%%%%%%%%%%%%%%%%%%%%%%%%%%%%%%%%%%%%%%%%%%%%%%%%%%%%
\section{Non-Gaussianity}\label{sec:bis}
Usually the 3-point correlator (bispectrum) is the first place to look for non-Gaussian signal. To calculate it, apart from the 3-point correlator of the noise, we also need to take into account the nonlinearity of the hydro plus inflaton equations \eqref{phieq} and \eqref{fleq}, which have to be expanded to second order in perturbations.
\comment{Using the definitions \eqref{def}, we find
\be\begin{split}
u^i &= \frac{1}{a^2} \d_i \Psi +\O(\E^3),\\[10pt]
\rho &= \rho_0 \left[1+ \E - \frac{4}{3 a^2} (\d_i \Psi)^2 +\O(\E^3)\right],\end{split}
\ee
where $\O(\E^3)$ is understood as cubic in any combination of $\Phi,\E,\Psi$.} A small subtlety is that the noise variance, being a local quantity, can no longer be assumed to be a constant $2\gamma_0 T_0$, but it is given by $2\gamma T\propto \rho$. Hence, a previously emitted long wavelength density perturbation will affect the subsequent fluctuations. To compute the bispectrum, an easy way to incorporate this effect is by defining
\be\label{xi2}
\xi = \sqrt{\frac{\rho}{\rho_0}} \xi_0 = \left(1+ \frac{\E}{2}+\O(\E^2)\right) \xi_0,
\ee
where the normalization of $\xi_0$ variance is $2\gamma_0 T_0$. Substituting \eqref{xi2} in \eqref{phieq} and \eqref{fleq}, neglecting slow-roll suppressed terms, using the background equations \eqref{sr} and dropping the index $0$ from background quantities, we obtain up to second order
\be\label{E2}\begin{split}
\dot \E + H \E + \frac{4}{3a^2 } \nabla^2 \Psi - 8 H \dot\Phi
& =  4 H \left[- \frac{3}{32}\E^2 + \frac{3}{2} \E \dot\Phi
+\dot\Phi^2+ \frac{1}{a^2} \d_i\Psi \d_i \Phi - \frac{1}{2 a^2 }(\d_j\Psi)^2\right]\\[10pt]
&~~~ +\frac{4 H }{\gamma \dot\phi} \left(1+ \dot\Phi + \frac{1}{2} \E\right) \xi_0 +\cdots\\[10pt]
\nabla^2(\dot \Psi+ 3H \Psi+ \frac{1}{4} \E+ 3 H\Phi)
&=-3 \d_i \Big[H\d_i \Phi \dot\Phi + \frac{3 H }{4} \E\d_i \Phi 
-\frac{1}{9 a^2}\d_i(\d_j \Psi)^2 + \frac{1}{3a^2} \d_j (\d_j \Psi \d_i \Psi) \\[10pt]
&~~~~~~~~~~~~+ \frac{H}{\gamma\dot\phi} \d_i \Phi \xi_0\Big] + \cdots\\[10pt]
 \ddot \Phi + (3H + \gamma) \dot\Phi+ \frac{3}{4} \gamma \E - \frac{1}{a^2} \nabla^2 \Phi
& =-\gamma\left[\frac{3}{4} \E \dot\Phi -\frac{3}{32} \E^2 - \frac{1}{2a^2}(\d_i\Psi)^2
+ \frac{1}{a^2}\d_i \Psi \d_i \Phi\right]\\[10 pt]
&~~~ - \frac{1}{\dot\phi} \left(1+ \frac{1}{2} \E\right)\xi_0 \cdots
\end{split}\ee
We did not switch to the conformal time because we use $t$ in our numerical integration.
%%%%%%%%%%%%%%%%%%%%%%%%%%%%%%%%%%%%%%%%%%%%%%%
\subsection{Estimates}\label{sec:est}
Before embarking on a detailed computation of the bispectrum, we use the structure of the second order equations, and what we have learned about the linear modes and the noise to make some general observations about non-Gaussianity in warm inflation.
%%%%%%%%%%%%%%%%%%%%%%%%%%%%%%%%%%%%%%%%%%
\subsubsection{Non-Gaussianity from nonlinear evolution}
It is common to define the following ratio as a measure of nonlinearity
\be\label{fnl}
f_{\rm NL}(k_1,k_2,k_3) = -\frac{5 \expect{\zeta_{\k_1}\zeta_{\k_2}\zeta_{\k_3}}'}{6 (P(k_1)P(k_2)+P(k_1)P(k_3)+P(k_2)P(k_3))}.
\ee
The RHS is indeed just a function of the moduli $k_1,k_2,k_3$: by translation invariance $\k_1,\k_2,\k_3$ form a triangle, and because of isotropy the orientation of the triangle is unimportant, so only its shape matters. If we focus on the contribution to the bispectrum coming from the nonlinearity of \eqref{E2}, the ratio \eqref{fnl} will be only a function of $\gamma/H$. To estimate it, we need to estimate the second order solution $\Phi^{(2)}$.

First consider the limit $\gamma\ll H$. We have $\E = \O\left(H^2 \Phi/\gamma\right)$ and therefore $\Phi^{(2)} = \O(H^2 \Phi^2/\gamma)$, which implies
\be
f_{\rm NL} \sim \frac{H}{\gamma},\qquad \text{when $\gamma \ll H$}.
\ee

\begin{figure}[t]
\centering
\includegraphics[scale = 1]{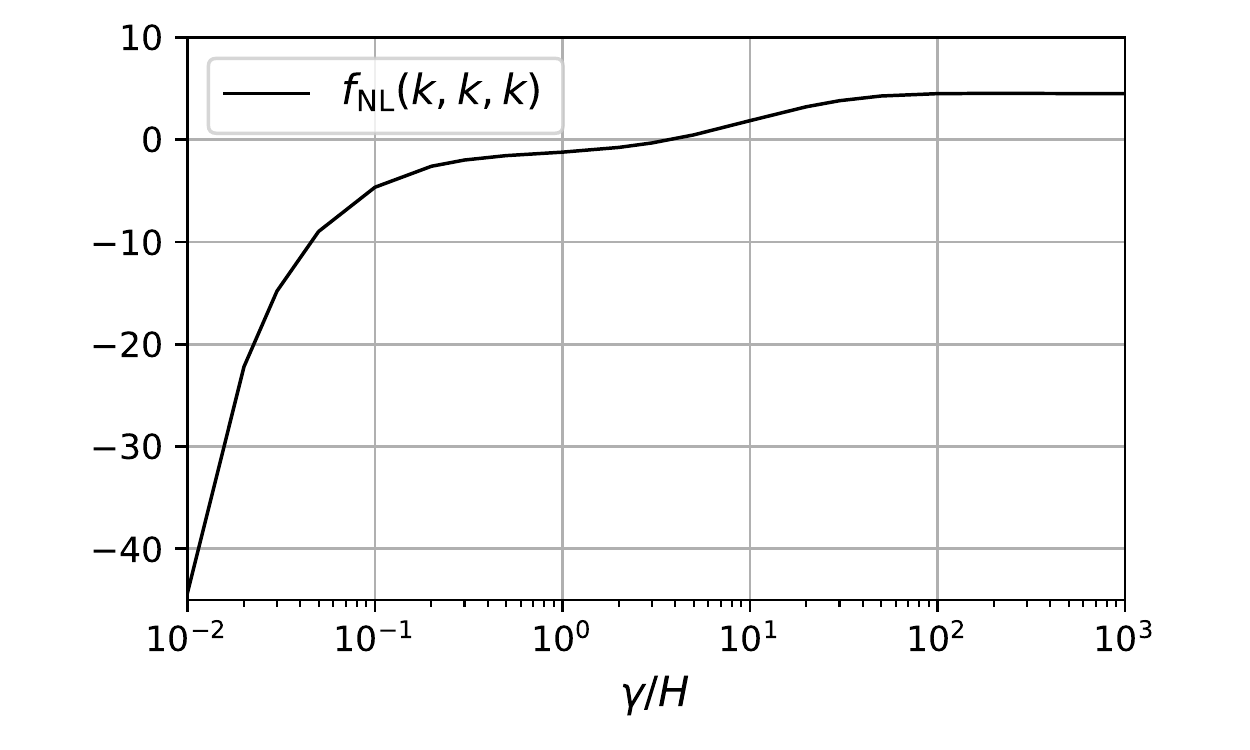} %[width= 11cm,height=8cm]                               
\caption{\small{Bispectrum from nonlinear evolution at the equilateral configuration and normalized as in \eqref{fnl}.}}
\label{fig:fnl}
\end{figure}

Next consider $\gamma\gg H$. Now we can focus on the slow mode that satisfies \eqref{slow}, and for which $\E= \O(H \Phi)$. This implies $\Phi^{(2)} = \O(H\Phi^2)$, or
\be
f_{\rm NL} \sim 1,\qquad \text{when $\gamma\gg H$}.
\ee
As an example, the function \eqref{fnl} evaluated at the equilateral configuration is plotted in figure \ref{fig:fnl}. At large $\gamma/H$ it asymptotes to $4.5$. (However, as can be seen from the figure \ref{fig:fs} the coefficient of the new-warm shape asymptotes much slower. Hence, the amplitude at the equilateral configuration is not a good measure of total non-Gaussianity in this regime.)
%%%%%%%%%%%%%%%%%%%%%%%%%%%%%%%%%%%%%%%%%%%%%%%%                                              
%%%%%%%%%%%%%%%%%%%%%%%%%%%%%%%%%%%%%%%%%%%%%%%           

On the other hand, we expect the squeezed-limit non-Gaussianity to vanish:
\be
\lim_{k_1\to 0} f_{\rm NL}(k_1,k_2,k_3)= 0.
\ee
The leading contribution in this limit comes from the interaction between a long (wavelength) mode and a short (wavelength) mode to generate a second order short mode. In the extreme limit, the long mode will be a combination of the superhorizon solutions \eqref{super}, which are all decaying except for the $\Phi$ and $\Psi$ components of $P_c$. These non-decaying components are constant plus $\O(k_1^2\eta^2)$ corrections. However, $\Phi$ and $\Psi$ never appear in the interactions without derivative acting on them, and hence the resulting second order solution vanishes when $k_1\to 0$. This is indeed expected from the fact that the background trajectory is an attractor. One can derive the squeezed-limit consistency condition, following the same argument as in \cite{Maldacena}, $\lim_{k_1\to 0}f_{\rm NL}(k_1,k_2,k_3)= \frac{5}{12} (n_s-1)$. Since $n_s-1$ is slow-roll suppressed, in our approximation the RHS vanishes.

However, the way this limit is reached in warm inflation is often distinct from single-field inflation, in which $f_{\rm NL}(k_1\ll k_2) \propto k_1^2/k_2^2$:

\begin{itemize}

  \item When $\gamma<H/5$ the decay of $P_-(\eta)\propto \eta^{\alpha_-}$ is slower than $\eta^2$, hence
    \be
    f_{\rm NL}(k_1\ll k_2) \propto \left(\frac{k_1}{k_2}\right)^{\alpha_-},\qquad \alpha_- = \frac{1}{2H}(4H+\gamma - \sqrt{4H^2 - 20 H\gamma +\gamma^2}).
    \ee
In particular, in the limit $\gamma\ll H$ we have $f_{\rm NL}(k_1\ll k_2) \propto k_1/k_2$. 
  
\item As discussed in section \ref{sec:asymp}, when $\gamma\gg H$ perturbations are dominated by emission at $-k\eta \sim \sqrt{\gamma/H}$ because afterward they grow until horizon-crossing at $-k\eta \sim 1$. Hence for moderate squeezing $\sqrt{H/\gamma} \ll \frac{k_1}{k_2}\ll 1$, most interaction terms between the long and short modes are as important as in the equilateral configuration, and $f_{\rm NL}$ will not decay.
  
  \item In this same regime, we can use the ``slow'' solution \eqref{fastslow} in the interaction term $\frac{1}{a^2} \d_i\Psi_{\rm long} \d_i \Phi_{\rm short}$ to infer that it is parametrically larger than the other interactions, 
\be\label{Bigint}
\frac{1}{a^2} \d_i\Psi_{\rm long} \d_i \Phi_{\rm short}\to  \frac{\k_1\cdot \k_2}{k_1^2} \Phi_{\k_1}\Phi_{\k_2}.
\ee
For symmetry reasons, the bispectrum is even in $\k_1\cdot\k_2$. Hence, from this enhanced interaction we obtain a non-decaying squeezed-limit $f_{\rm NL}$ with quadrupolar modulation. In summary, 
    \be
    f_{\rm NL} = f_0(k_1/k_2) + f_2(k_1/k_2)\ (\hat\k_1\cdot\hat\k_2)^2,\qquad \sqrt{H/\gamma}\ll \frac{k_1}{k_2}\ll 1
    \ee
    where $f_0$ and $f_2$ are $\O(1)$, slowly varying functions. In the limit $\gamma\to \infty$, when the ``slow'' mode has exact scaling behavior in the range $1\ll -k_1\eta\ll \sqrt{\gamma/H}$, $f_0$ and $f_2$ become constants. Given that in most cosmological measurements of the bispectrum, $k_1/k_2$ cannot be made arbitrarily small, in practice a combination of the local template and its quadrupole version would perhaps be a good fitting function in the limit of extremely large $\gamma/H$. 
  \item The enhanced interaction \eqref{Bigint} does lead to a growth $\propto k_2/k_1$ in the 4th and higher order squeezed correlators. But it requires a very large squeezing (and hence very large $\gamma/H$) to overcome the overall suppression of the signal in these correlators compared to the bispectrum.
\end{itemize}
Another feature of the bispectrum in the $\gamma\gg H$ regime is the enhancement of the folded triangles, e.g. $k_3 = k_1+k_2$. Consider $k_1 = k_2 = k_3/2$. Then the interaction between $k_1$ and $k_2$ generates a shorter mode $k_3$, which grows for a longer time after $k_1$ and $k_2$ freeze. The horizon crossing times are related by $\eta_1 = 2\eta_3$. In the limiting case, where the growth of the slow mode is proportional to $\eta^{-9}$, this gives a factor of $2^9$. 
%%%%%%%%%%%%%%%%%%%%%%%%%%%%%%%%%%%%%%%
\subsubsection{Shot-noise non-Gaussianity}
Let us now return to the non-Gaussianities arising directly from the source. In the hydro description, we can use \eqref{xiN} to obtain
\be\label{zNPoiss}
\expect{\zeta_{\k_1}\cdots \zeta_{\k_N}}'= \Gamma\ \lambda_N\left(-\frac{H}{f\dot\phi}\right)^N \int_{-\infty}^0 d\eta \  (H\eta)^{4(N-1)} \prod_{i=1}^N G_{\Phi_{\k_i}}(0,\eta),
\ee
where $\lambda_N=1$ for even $N$ and $\lambda_N=\dot\phi/(2f T)\ll 1$ for odd $N$. If this integral converges, then it must be dominated by scales of order $H$ (for simplicity, we assume $\gamma$ is not much bigger). Therefore, we get a simple estimate of non-Gaussianity 
\be\label{zN/zrms}
\frac{k^{3(N-1)}\expect{\zeta^N}'}{\zrms^N} \sim \lambda_N \left(\frac{H^4}{\Gamma}\right)^{(N-2)/2}.
\ee
This is small as a manifestation of the central limit theorem. There are $\O(\Gamma/H^4)$ independent fully non-Gaussian contributions to the Hubble scale fluctuations. They average to an approximately Gaussian distribution. As discussed earlier, we do not expect $\Gamma$ to be ever much larger than $T^4$, though at small coupling it is natural to expect $\Gamma \ll T^4$.

However, a closer look at the subhoriozn behavior of the Green's functions in \eqref{zNPoiss} reveals that for any $\gamma/H$, there is a large enough $N$ such that the integral diverges as $\eta\to -\infty$. This is the ultraviolet limit, and the divergence is an artifact of replacing the short-range correlated noise by an exactly delta-function correlated one. In reality, the integrals would be cut off once $|k\eta|>1/H\ell$, and we get a finite answer, but one that depends on the microscopic details via some ``form factors'' characterizing the shape of the noise correlators. 

Hence there is a source of UV sensitivity in our effective description, encoded in the non-Gaussian initial state of fluctuations. We can nevertheless use \eqref{zNPoiss} with a cutoff to estimate this effect. The decay of Green's functions at large $|\eta|$ is slowest when $\gamma\ll H$, in which case all $N\geq 4$ integrals have to be cut off. In this limit, $G_\Phi(0,\eta) \propto \eta^{-3}$, and cutting off the integral changes the estimate \eqref{zN/zrms} to
\be
\frac{k^{3(N-1)}\expect{\zeta^N}'}{\zrms^N} \sim \lambda_N \frac{H^2}{\Gamma^{1/2}} \left(\frac{H}{\Gamma^{1/2} \ell}\right)^{N-3}.
\ee
As explained in \cite{Babich}, the ratio on the left determines how visible the $N$th order non-Gaussianity is. Taking $T$ to be the only scale in the problem, we see that it still decays with $N$.\footnote{In principle, if the dissipative process is so rare and localized that $\Gamma \ell^2 < H^2$ the ratios grow with $N$. (In the case of sphaleron heating, this cannot happen as we will shortly see.) Then the lowest order correlators will not necessarily be the most visible non-Gaussianity. As a simple example consider the distribution
  \be
  p(x) =A e^{-x^2/2} +\ep \delta (x- x_0),\qquad \ep \ll 1.
  \ee
  The significance of the non-Gaussian moments at $\O(\ep)$ is determined by
  \be
  \frac{\expect{x^N}_\ep}{\sqrt{\expect{x^{2N}}_0}} = \ep \sqrt{\frac{N!}{(2N)!}} 2^{N/2} x_0^N.
  \ee
  For $x_0\gg 1$, this is peaked at $N\sim x_0^2$. }
In fact, despite the extra suppression by $\dot\phi/(2fT)$, the bispectrum is still dominant compared to the trispectrum (4-point correlator).

Even though strictly speaking there is no divergence at $N=3$, the integral \eqref{zNPoiss} converges very slowly in the folded limit $k_3 = k_1 + k_2$, going as $\int_{-\infty} d\eta/\eta^{1+3\gamma/2H}\sim H/\gamma$. Therefore, the signal in the this limit is expected to be enhanced by ${\rm min}(H/\gamma, -\log(H\ell))$. Away from the exact limit this enhancement would be saturated by $\log\left(k_1+k_2-k_3\right)$, and therefore it is challenging to see it in the data. Numerically, we hardly see any sign of this logarithmic growth.

One should ask if the shot-noise non-Gaussianity can ever be comparable with the other kind. Naively, it is enough to reduce the YM coupling to decrease $\Gamma$ and scale up \eqref{zN/zrms}. However, to keep the scalar power noise-dominated, $\alpha$ and $f$ have to be sent to zero together. Soon $\kappa = \alpha\dot\phi/2\pi f$ (which is the momentum scale that our coupling introduces in the gauge field equation of motion) grows above the microscopic scale of relevance, namely the sphaleron scale $\alpha T$. At this point our analysis based on a the linear response is no longer valid. To avoid this, we require
\be\label{alpha}
\frac{\dot\phi}{2 \pi  T f}<1\Rightarrow \sqrt{\frac{2 H \rho}{\pi^2 \alpha^2 T \Gamma}} <1,
\ee
where we used \eqref{sr} and \eqref{FDT} to obtain the second expression. Since $\Gamma\sim \alpha^5 T^4$, this condition does not let $\alpha$ to be very small. Indeed, we can use \eqref{T(Y)} to find a lower bound on $\alpha$ for a given $\gamma/H$ and $g_*$. This would imply $f_{\rm NL}^{\rm shot-noise}< H/\gamma$, which is subleading compared to what we get from nonlinear evolution. We will see an explicit example below.
%%%%%%%%%%%%%%%%%%%%%%%%%%%%%%%%%%%%%%
\subsubsection{Gravitational interactions}
We have neglected gravitational interactions because at the level of the bispectrum they lead to $f_{NL} \sim \ep$ as in \cite{Maldacena}, while we are finding $f_{NL} = \O(1)$. Generally, gravitational contribution to non-Gaussianities is slow-roll suppressed compared to those coming from inflaton-phonon interactions. This can be understood by using the Einstein equation to estimate metric perturbations
\be
h\sim \frac{\dot\phi^2}{\mpl^2} \Phi^2.
\ee
Substituting this in the inflaton and hydro equations and estimating derivatives with $H$, gives interaction vertices in \eqref{E2} that are of order
\be
\frac{\dot\phi^2}{\mpl^2 H^2}\sim \ep.
\ee
%%%%%%%%%%%%%%%%%%%%%%%%%%%%%%%%%%%%%%%
\subsubsection{Loops}
Because of the subhorizon excitations, loop corrections in warm inflation are qualitatively different from those in cold inflation. Nevertheless, they introduce small corrections. Firstly, since superhorizon adiabatic perturbations are locally unobservable, there is negligible contribution to their correlators after horizon crossing. So we need to investigate contributions from the time of horizon crossing and before. Loop contributions from the modes with similar wavelength as the ones of interest, which therefore cross the horizon around the same time, is controlled by the variance $\zeta_{\rm rms}^2\ll 1$. There can also be sizable UV contributions to the loop integrals. These loops include thermal excitations and therefore their effect is more subtle and might in principle be important for some range of parameters. We leave a careful analysis of loops to future work. Still, our restriction to the tree level is a formally well-defined truncation.

%%%%%%%%%%%%%%%%%%%%%%%%%%%%%%%%%%%%
\subsubsection{Parity violation}
One last feature, specific to the microscopic axionic model, that we would like to comment on is parity violation. We should expect this to enter at some level in the cosmological predictions. As far as scalar correlators are concerned, the first indication of parity violation would be the 4-point function; bispectrum is parity invariant because the mirror image of a triangle in 3-dimensions can be related to itself by a rotation, which is a symmetry of the model. In our effective description, parity violation can enter as field-dependent corrections to the noise correlators, and also as higher derivative corrections to the effective hydro description. For instance, the four-velocity defined by the energy-momentum tensor, might not coincide with the one entering the expectation value of $\Tr[G \tilde G]$:
\be
\expect{\frac{\alpha}{16 \pi f}\Tr[G_{\mu\nu}\tilde G^{\mu\nu}]}= 
\gamma \left(u^\mu +\frac{1}{M} \vep^{\mu\nu\alpha\beta}u_\nu D_\alpha u_\beta\right) \d_\mu \phi,
\ee
for some $M \sim T$. This results in a parity-violating trispectrum with $g_{NL} \sim H/T$, which is unfortunately too small to be observable in the near future. See \cite{Liu} for another inflationary model with parity violating trispectrum.

%%%%%%%%%%%%%%%%%%%%%%%%%%%%%%%%%%%%%%%%%%%%%%%%%%%%
\subsection{Shapes and templates}\label{sec:shapes}
In our approximation, the bispectrum is divided into two pieces
\be
\expect{\zeta_{\k_1}\zeta_{\k_2}\zeta_{\k_3}}'=B_{211}(k_1,k_2,k_3)+B_{\rm sn}(k_1,k_2,k_3).
\ee
The first piece originates from the nonlinear evolution 
\be
B_{211}(k_1,k_2,k_3)=-H^3 \expect{\Phi^{(2)}_{\k_1} \Phi^{(1)}_{\k_2}\Phi^{(1)}_{\k_3}}'+ \text{2 cycl.}
\ee
where 2 cycl. correspond to terms with $k_1\leftrightarrow k_2$ and $k_1\leftrightarrow k_3$, and $\Phi^{(2)}$ is the second order solution from \eqref{E2}. The second piece is the shot noise, non-Gaussianity from the source:
\be
B_{\rm sn}(k_1,k_2,k_3)= -H^3 \expect{\Phi^{(1)}_{\k_1} \Phi^{(1)}_{\k_2}\Phi^{(1)}_{\k_3}}'.
\ee
To explain how these correlators are computed in practice, it is useful to consider a toy model
\be
L\Phi  = \frac{g}{a(t)^2} |\nabla\Phi|^2+ \xi,
\ee
where $L$ is a linear differential operator, e.g.
\be
L\Phi = \ddot\Phi + 3H \dot\Phi - \frac{1}{a(t)^2}\nabla^2\Phi,
\ee
and 
\be
\expect{\xi_{\k}(t_1) \xi_{-\k}(t_2)}' = \frac{A_2}{a(t_1)^3} \delta(t_1-t_2),
\ee
\be
\expect{\xi_{\k_1}(t_1) \xi_{\k_2}(t_2) \xi_{\k_3}(t_3)}' = \frac{A_3}{a(t_1)^6} \delta(t_1-t_2) \delta(t_1-t_3).
\ee
Denote by $\hat G_k (t_1,t_2)$ the retarded Green's function of $L$ in momentum space. Then 
\be
P(k) = H^2 \lim_{t\to \infty} \expect{\Phi^{(1)}_\k(t)\Phi^{(1)}_{-\k}(t)}' 
= A_2 H^2 \int_{-\infty}^{\infty} \frac{dt}{a(t)^3} \hat G_{k}(\infty,t)^2,
\ee
\be
B_{\rm sn}(k_1,k_2,k_3) = 
-A_3 H^3\int_{-\infty}^{\infty} \frac{dt}{a(t)^6} \hat G_{k_1}(\infty,t) \hat G_{k_2}(\infty,t) \hat G_{k_3}(\infty,t),
\ee
and 
\be\begin{split}
B_{211}(k_1,k_2,k_3) = &
2g  A_2^2 H^3\ \k_2\cdot \k_3\int_{-\infty}^{\infty} \frac{dt_1}{a(t_1)^2} \hat G_{k_1}(\infty,t_1)
\int_{-\infty}^{t_1} \frac{dt_2}{a(t_2)^3} \hat G_{k_2}(\infty,t_2)\\[10pt]
&~~~~~~~~~~~~~\times\int_{-\infty}^{t_1} \frac{dt_3}{a(t_3)^3} \hat G_{k_3}(\infty,t_3)\ \ + \text{2 cycl.}
\end{split}\ee
In our actual model, the Green's functions and the analogs of the above integrals are computed numerically. As written, $B_{211}$ is a three-dimensional time integral, which is computationally very expensive. However, it can be evaluated much faster using a trick from \cite{Bastero-Gil}. Let us define
\be
p(k;t) = H^2 \expect{\Phi_{\k}^{(1)}(t) \Phi_{-\k}^{(1)}(\infty)}',
\ee
and
\be
b(k_1,k_2,k_3;t) =- H^3\expect{\Phi_{\k_1}^{(2)}(t) \Phi_{\k_2}^{(1)}(\infty)\Phi_{\k_3}^{(1)}(\infty)}',
\ee
in terms of which
\be
P(k) = p(k;\infty)
\ee
and
\be
B_{211}(k_1,k_2,k_3)= b(k_1,k_2,k_3;\infty)+b(k_2,k_3,k_1;\infty)+b(k_3,k_1,k_2;\infty).
\ee
These satisfy
\be\label{LF}
L_\k p(k;t) =A_2H^2 \frac{\hat G_k(\infty,t)}{a(t)^3},
\ee
and
\be\label{LB}
L_{\k_1} b(k_1,k_2,k_3;t) = 2 \frac{g}{H a(t)^2} (\k_2\cdot\k_3) p(k_2;t) p(k_3;t).
\ee
Hence, by first tabulating $\hat G_{k}(\infty,t)$, we can numerically integrate the deterministic system \eqref{LF} and \eqref{LB}. The computational cost of this method is a one (rather than three) dimensional time integral.

\begin{figure}[t]
\centering
\includegraphics[scale =0.7]{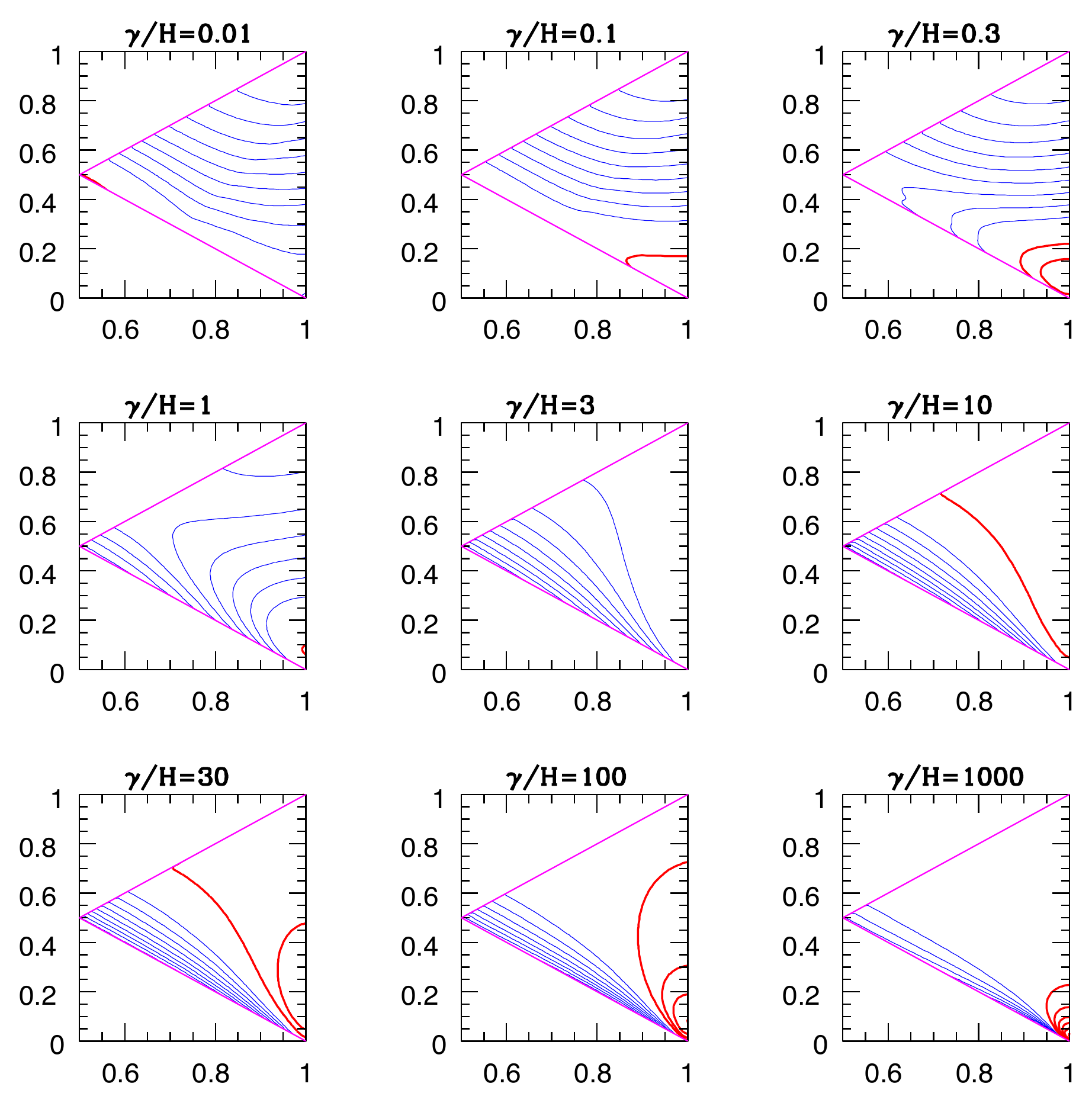} %[width= 11cm,height=8cm] 
\caption{\small{
Mosaic of $x_2^2 x_3^2 B_{211}(1,x_2,x_3)$. Positive (thick red) and negative (thin blue) multiples of $0.1 {\rm max}(|x_2^2 x_3^2 B_{211}(1,x_2,x_3)|)$.}}
\label{fig:B211}
\end{figure}

\begin{figure}[t]
\centering
\includegraphics[scale =0.7]{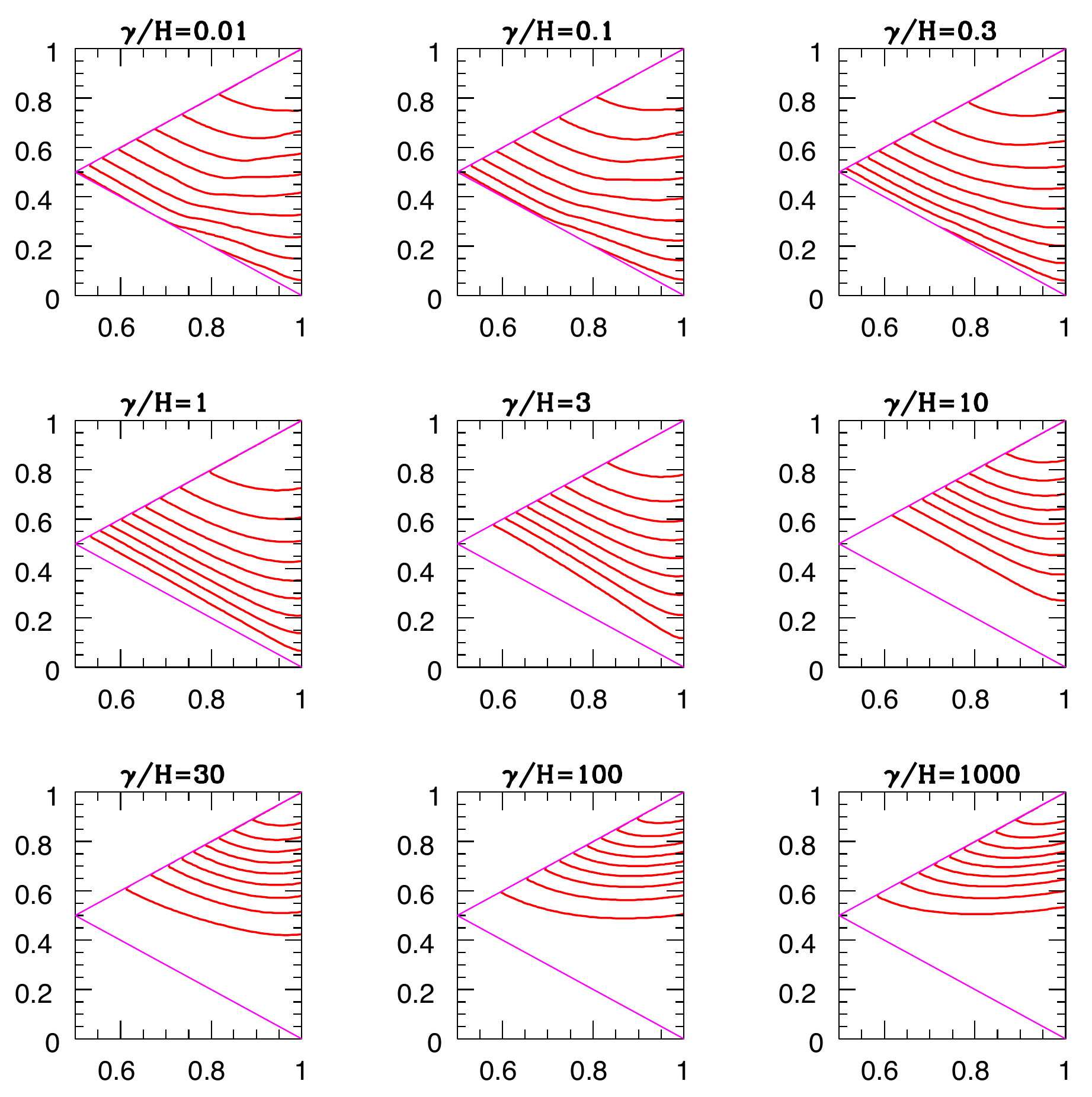} %[width= 11cm,height=8cm]                               
\caption{\small{Mosaic of $x_2^2 x_3^2 B_{\rm sn}(1,x_2,x_3)$. Positive (thick red) and negative (thin blue) multiples of $0.1 {\rm max}(|x_2^2 x_3^2 B_{\rm sn}(1,x_2,x_3)|)$.}}
\label{fig:Bst}
\end{figure}

We apply the same method to integrate \eqref{E2}. The resulting contours of $B_{211}$ and $B_{\rm sn}$ for several choices of $\gamma/H$ are shown respectively in figures \ref{fig:B211} and \ref{fig:Bst}. To characterize the shape, it is common to use the scale-invariance (up to slow-roll corrections) to write
%%%%%%%%%%%%%%%%%%%%%%%%%%%%%%%%%%%%%%%%%%%%%%%%                                              
%%%%%%%%%%%%%%%%%%%%%%%%%%%%%%%%%%%%%%%%%%%%%%%           
\be
B(k_1,k_2,k_3) = \frac{1}{k_1^6} B(1,x_2, x_3),\qquad x_{2,3} \equiv \frac{k_{2,3}}{k_1},
\ee
then use invariance under permutations to take $k_1$ to be the largest momentum, and $x_3<x_2<1$. Triangle inequality implies $1-x_3\leq x_2$. The contour plots show $x_2^2 x_3^2 B(1,x_2,x_3)$, because the appropriate inner product between two shapes in a perfect $3d$ cosmic variance dominated survey is \cite{Babich}
\be\label{dot}
B_1 \cdot B_2 = \int_{1/2}^1 dx_2 \int_{1-x_2}^{x_2}\!dx_3 x_2^4 x_3^4 B_1(1,x_2,x_3) B_2(1,x_2,x_3).
\ee
%%%%%%%%%%%%%%%%%%%%%%%%%%%%%%%%%%%%%%%%%%%%%%%%                                              
%%%%%%%%%%%%%%%%%%%%%%%%%%%%%%%%%%%%%%%%%%%%%%%           

Since the noise was assumed to be Gaussian in \cite{Bastero-Gil}, there is no analog of $B_{\rm sn}$ to compare our result with. (This type of non-Gaussianity  is often encountered in particle production scenarios e.g. \cite{GW,Flauger}, and a particular example has been constrained by the CMB data \cite{Smith}.) However, $B_{211}$ can be compared with \cite{Bastero-Gil}, and we find qualitative but not quantitative agreement. For instance, the plot of the $f_{\rm NL}$ function in figure \ref{fig:fnl} shows a change of sign unlike the FIG. 1 of \cite{Bastero-Gil}. As a check of our system of second order perturbations, we have performed a $1+1 d$ simulation of the exact non-linear equations \eqref{phieq} and \eqref{fleq} and have confirmed that turning on a small, slowly varying source $\xi = \ep_1(t) \sin(k_1 x)+\ep_2(t) \sin(k_2 x)$ produces (within our numerical resolution) the same second order perturbation predicted by \eqref{E2}. 

\begin{figure}[t]
\centering
\includegraphics[scale =1]{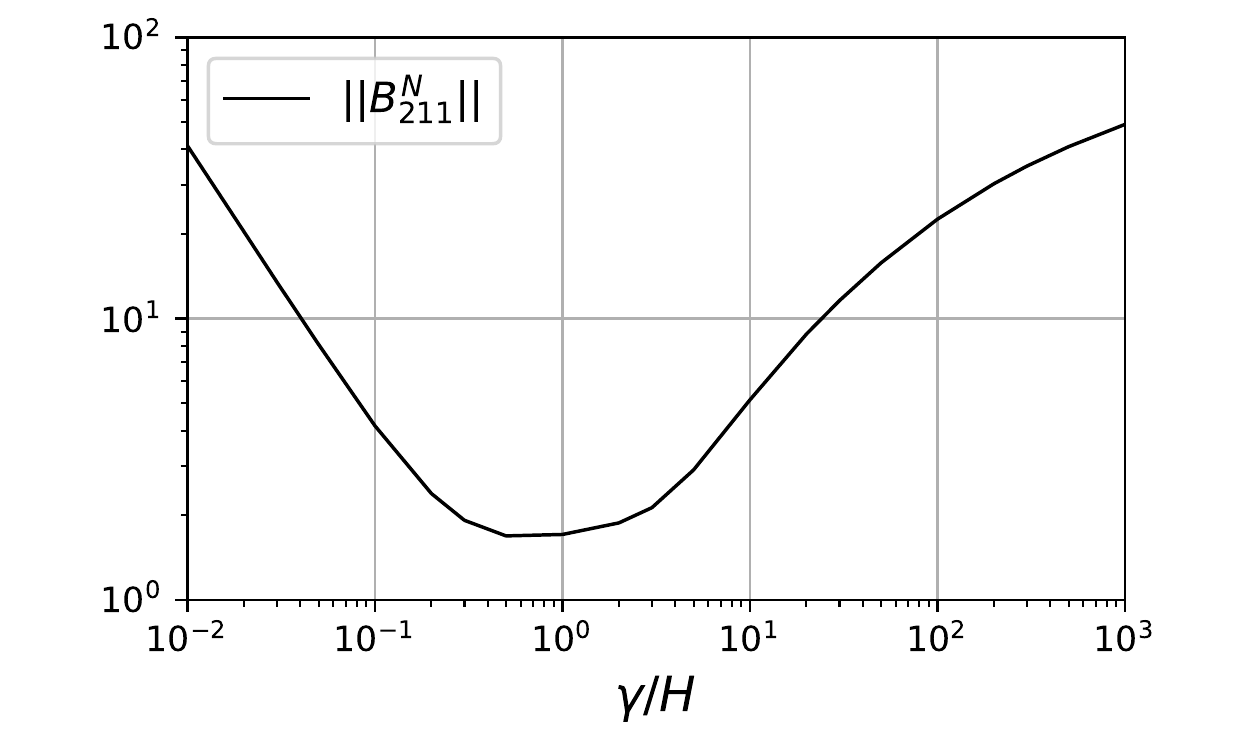} %[width= 11cm,height=8cm] 
\caption{\small{The norm of the $211$ bispectrum of the Newtonian potential $-3 \zeta/5$.}}
\label{fig:norm}
\end{figure}

In figure \ref{fig:norm}, we show the norm of the $211$ bispectrum expressed for the Newtonian potential $\Phi_N = -3 \zeta/5$
\be
||B^N_{211}|| \equiv \left(\frac{3}{5}\right)^3 \sqrt{B_{211}\cdot B_{211}}.
\ee
This shows that despite the smallness of $f_{\rm NL}(k,k,k)$ around $\gamma\approx H$, the full bispectrum remains nontrivial in the entire range of $\gamma/H$. 

%%%%%%%%%%%%%%%%%%%%%%%%%%%%%%%%%%%%%%%%%%%%%%%%                                              
%%%%%%%%%%%%%%%%%%%%%%%%%%%%%%%%%%%%%%%%%%%%%%%   

The bispectrum as a nontrivial function of two variables contains a lot of information, but for an initial search or putting constraints, it is enough to analyze the data against a simpler template that is well correlated with the model prediction. In particular, it is useful to have an analytic expression that can be written as a sum of factorizable terms. We propose one new template, the ``new-warm'' shape, in addition to the equilateral template for this purpose. Let us introduce the building blocks
\be
F_{ab} = A^2 k_1^{-a} k_2^{-b} k_3^{-6+a + b} + \text{5 perms.}
\ee
where $A=9 \zrms^2/25$ is the normalization of the power spectrum for the Newtonian potential. Then, with correlation measured using
\be
C(B_1,B_2) \equiv \frac{B_1\cdot B_2}{\sqrt{(B_1\cdot B_1)(B_2\cdot B_2)}},
\ee
and within the range of $\gamma/H$ that we have studied
\begin{itemize}
\item $B_{\rm sn}$ is always highly correlated with the standard equilateral template \cite{Creminelli_eq}:
\be
F^{\rm eq} = 6 F_{32} - 3 F_{33} -2 F_{22}.
\ee
\item $B_{211}$ is also highly correlated with the equilateral template for $\gamma \leq H$.
\item $B_{211}$ is highly correlated with the new-warm template for $\gamma \geq H$:
  \be
F^{\rm nw} = 2 F_{43} +F_{42}- F_{44}- F_{33}-F_{41} +0.08 F^{\rm eq},
\ee
which is orthogonal to the equilateral template.
\end{itemize}
%%%%%%%%%%%%%%%%%%%%%%%%%%%%%%%%%%%%%%%%%%%%%%%%                                              
\begin{figure}[t]
\centering
\includegraphics[scale =1]{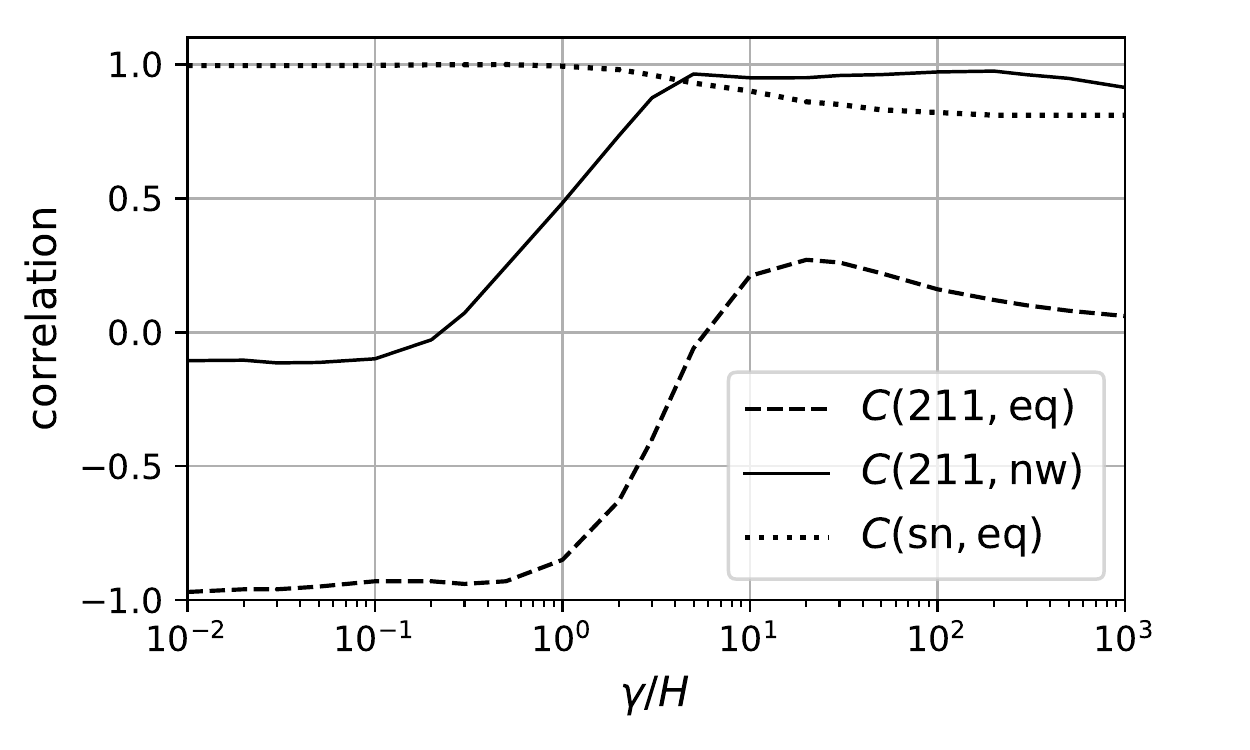} %[width= 11cm,height=8cm] 
\caption{\small{The correlation of the $211$ bispectrum with the equilateral and new-warm templates.}}
\label{fig:cor}
\end{figure}
%%%%%%%%%%%%%%%%%%%%%%%%%%%%%%%%%%%%%%%%%%%%%%%           

These correlations are shown in figure \ref{fig:cor}. The new template $F^{\rm nw}$ is about $60\%$ correlated with the $B_S$ template introduced in \cite{Bastero-Gil} in order to cover the large $\gamma/H$ regime. Despite the quantitative mismatches, this does look similar to our bispectrum shape at large $\gamma/H$. Our main motivation for introducing $F^{\rm nw}$ is that $B_S$ is defined via a cutoff $k_3> 0.1 k_1$, which breaks factorizablility.

Figure \ref{fig:fs} shows the $f_{\rm NL}$ coefficients for the 211 non-Gaussianity, as conventionally defined in terms of the Newtonian potential. They are fixed in terms of $\gamma/H$. The strongest constraint on the equilateral shape is \cite{Planck}
\be
f_{\rm NL}^{eq} = -26\pm 47.
\ee
This could already rule out the range $\gamma/H<10^{-3}$. However, because of \eqref{gamma_min}, we cannot make $\gamma/H$ arbitrarily small while having a noise-dominated power. On the other hand, if the vacuum contribution dominates in \eqref{zxi+vac}, the signal becomes more Gaussian than what we are presenting. It would be interesting to also search for the new-warm template. 

\begin{figure}[t]
\centering
\includegraphics[scale =1.]{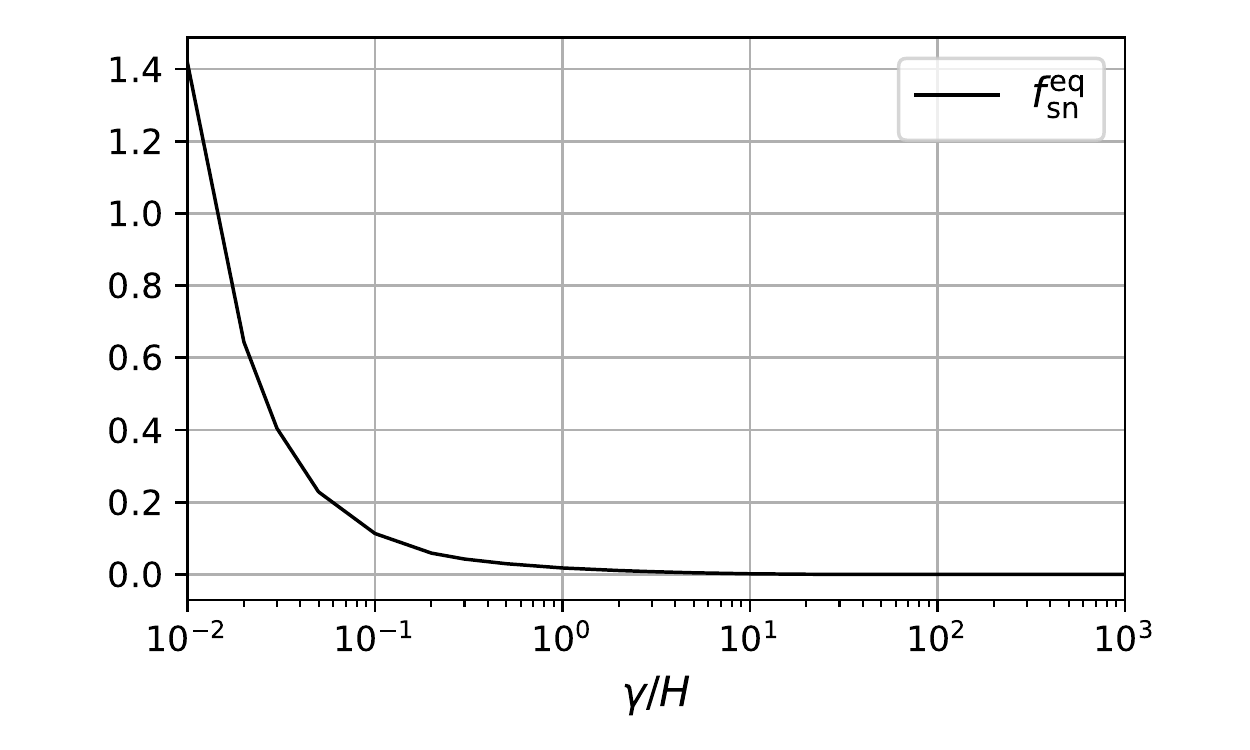} %[width= 11cm,height=8cm] 
\caption{\small{The maximum amplitude of equilateral template in the shot-noise non-Gaussianity, assuming inflaton is coupled to $SU(2)$ YM.}}
\label{fig:fsneq}
\end{figure}

The strength of $B_{\rm sn}$ is not fixed by $\gamma/H$, so there is no unique prediction for $f_{\rm NL}^{\rm eq}$. However, it would be useful to focus on a particular case and obtain a rough estimate of the numerical value. An explicit expression for $\Gamma$ in $SU(N)$ theory in terms of $\alpha$ and $T$ has been proposed in \cite{Moore}:
\be\label{Gamma}
\Gamma \approx h_* \alpha^5 T^4,
\ee 
where $h_*$ depends on $N$ and logarithmically on $\alpha$. For instance, $h_* \approx 134$ for $SU(2)$ gauge group at $\alpha = \frac{1}{2}$. Assuming a noise dominated power spectrum, $T/H$ can be determined as a function of $\gamma/H$ using \eqref{T(Y)}, and we can eliminate $\frac{H}{f\dot\phi}$ in terms of $\Gamma$, $F_2(\gamma/H)$ and $\zrms$. Using \eqref{sr} and \eqref{FDT}, we find
\be
\expect{\zeta_{\k_1}\zeta_{\k_2}\zeta_{\k_3}}' 
= - \sqrt{\frac{2H \rho}{T\Gamma^2}} \left(\frac{\zrms^2}{F_2(\gamma/H)}\right)^{3/2}
\int_{-\infty}^\infty \frac{dt}{a(t)^6} G_{\Phi_{\k_1}}(\infty,t)G_{\Phi_{\k_2}}(\infty,t)G_{\Phi_{\k_3}}(\infty,t).
\ee
To get the maximum possible value consistent with our assumptions, we find the smallest $\alpha$ that is compatible with the limit in \eqref{alpha}:
\be 
{\rm min}(\alpha) \approx \left(\frac{\pi^2 g_*^4 \zrms^2 H^2}{50625 h_*^3 F_2(\gamma/H) \gamma^2}\right)^{1/15}.
\ee
The coefficient of the equilateral shape for this choice is plotted in figure \ref{fig:fsneq}. As anticipated this is smaller than $f_{211}^{\rm eq}$.  

%%%%%%%%%%%%%%%%%%%%%%%%%%%%%%%%%%%%%%%%%%%%%%%%                                              
%%%%%%%%%%%%%%%%%%%%%%%%%%%%%%%%%%%%%%%%%%%%%%%           

%%%%%%%%%%%%%%%%%%%%%%
\section{Discussion}\label{sec:con}
%%%%%%%%%%%%%%%%%%%%%
%%%%%%%%%%%%%%%%%%%%%%
Minimal warm inflation appears to be a successful realization of particle production and thermalization during inflation. One that is still consistent with observational constraints, and potentially distinguishable from single-field inflation. We have seen that it is a local attractor. We do not know how large the basin of attraction is, and where is the boundary with other scenarios with nontrivial background gauge fields \cite{Shahin,Adshead,Kamali}. 

Of course, there are other inflationary models with sizable particle production e.g. \cite{Anber,trapped,GW,Flauger,Abolhasani}, and these all share the feature of introducing extra sources for primordial fluctuations. They often predict sizable (or even too large) non-Gaussianity. However, unlike minimal warm inflation they operate at zero temperature and take advantage of an instability that is triggered by the rolling of the inflaton field, and regulated by the Hubble expansion. As such, they would lead to an exponential runaway on a flat metric, as opposed to the milder algebraic heating by sphalerons. It is worth exploring the possibility of thermalization in such models \cite{Ferreira}. We expect them to work in qualitatively different ways at temperatures above the instability scale.

As an example, let us consider replacing the YM theory with QED. For $\alpha\to 0$ with $\alpha/f$ kept fixed, this is the model of \cite{Anber} where there is an instability and exponential sensitivity to $\dot\phi$. For larger $\alpha$, and small enough $m_e$, it has been argued in \cite{Domcke} that Schwinger pair production leads to a milder power-law dependence. There the focus was on the regime where the produced particles do not thermalize. What about the thermal regime?

Let us define $Q_A(t)$, the analog of the random-walking variable in YM, by replacing $\Tr[G\tilde G] \to F\tilde F$ in \eqref{Q}. It can be written as $Q_A(t) =\frac{\alpha}{4\pi}(\H(t) - \H(0))$, where
\be
\H = \int_V d^3 \r A\cdot B
\ee
is called the ``magnetic helicity''. In contrast to YM, this is gauge invariant and no longer periodic. Hence, there are no sphalerons at finite temperature and if we put the system in a box of size $L \ll f/\dot\phi$ not much will happen.\footnote{One can instead build a nice example of sphaleron heating in two-dimensional Abelian Higgs model \cite{Rubakov}.} 

For large $L$, the coupling $\phi F\tilde F$ does have an effect because it induces an instability of the modes with momentum of order or less than $\kappa \equiv \alpha\dot\phi/(2\pi f)$. The flux of energy into the plasma is
\be\label{dotrho}
\dot\rho = \kappa E\cdot B. 
\ee
This can be seen from the fact that at constant $\kappa$, there is a conserved Hamiltonian
\be
H = H_{\rm QED} + \frac{1}{2}\kappa \H.
\ee
To estimate $\expect{E\cdot B}$, suppose $T\gg m_e$ which implies the conductivity of the plasma is $\sigma \sim T\gg \kappa$. In this case, we can estimate
\be\label{E}
E\sim \frac{\kappa}{T} B,\qquad \text{momentum $\sim \kappa$.}
\ee
There is little contribution to $\expect{E\cdot B}$ from higher momenta. So we need to find the magnetic field at the instability scale. This can be determined using the fact that if we inject a large amount of energy at scale $\kappa$, it launches a turbulent cascade to the shorter scales where it dissipates into heat. The quasi-stationary state of the system is determined by equating the cascade rate with the input rate. The cascade rate at scale $\lambda$ is 
\be
\frac{\rho v_\lambda^2}{\lambda/v_\lambda} \sim \frac{B^3_\lambda}{\lambda T^2}
\ee
where we estimated $v$ by the Alfv\'en speed $\sqrt{B^2_\lambda/\rho} \sim B_\lambda/T^2$. Comparison with \eqref{dotrho} and \eqref{E} gives $B_\kappa \sim \kappa T$ and
\be
\dot \rho \sim \kappa^4 T.
\ee
This rate is much smaller than the sphaleron heating $\propto \kappa^2 T^3$. Perhaps this heating mechanism could still sustain a hot plasma during inflation if $\kappa \gg H$, but there are qualitative differences compared to the non-Abelian case. For instance, since the heating rate is proportional to $\dot \H$, helical magnetic fields are constantly generated and inverse-cascade to larger scales. This is likely to result in too large non-Gaussianity.

So far, heating up inflation via sphalerons looks not only minimal but also quite exceptional. 
\vspace{0.3cm}
\noindent
\section*{Acknowledgments}

We thank Paolo Creminelli, Sergei Dubovsky, Arkady Vainshtein, Alex Vilenkin, Matias Zaldarriaga and especially Giovanni Villadoro for useful discussions. We also thank William DeRocco, Peter Graham, Saarik Kalia, and Mikko Laine for comments on the first version. 

\bibliography{bibwng}

\providecommand{\href}[2]{#2}\begingroup\raggedright\begin{thebibliography}{10}

\bibitem{Moss}
I.~G. Moss, ``{Primordial Inflation With Spontaneous Symmetry Breaking},''
  \href{http://dx.doi.org/10.1016/0370-2693(85)90570-2}{{\em Phys. Lett. B}
  {\bfseries 154} (1985) 120--124}.

\bibitem{Maeda}
J.~Yokoyama and K.-i. Maeda, ``{On the Dynamics of the Power Law Inflation Due
  to an Exponential Potential},''
  \href{http://dx.doi.org/10.1016/0370-2693(88)90880-5}{{\em Phys. Lett. B}
  {\bfseries 207} (1988) 31--35}.

\bibitem{Fang}
A.~Berera and L.-Z. Fang, ``{Thermally induced density perturbations in the
  inflation era},'' \href{http://dx.doi.org/10.1103/PhysRevLett.74.1912}{{\em
  Phys. Rev. Lett.} {\bfseries 74} (1995) 1912--1915},
  \href{http://arxiv.org/abs/astro-ph/9501024}{{\ttfamily
  arXiv:astro-ph/9501024}}.

\bibitem{Graham_09}
C.~Graham and I.~G. Moss, ``{Density fluctuations from warm inflation},''
  \href{http://dx.doi.org/10.1088/1475-7516/2009/07/013}{{\em JCAP} {\bfseries
  07} (2009) 013}, \href{http://arxiv.org/abs/0905.3500}{{\ttfamily
  arXiv:0905.3500 [astro-ph.CO]}}.

\bibitem{Bastero-Gil_11}
M.~Bastero-Gil, A.~Berera, and R.~O. Ramos, ``{Shear viscous effects on the
  primordial power spectrum from warm inflation},''
  \href{http://dx.doi.org/10.1088/1475-7516/2011/07/030}{{\em JCAP} {\bfseries
  07} (2011) 030}, \href{http://arxiv.org/abs/1106.0701}{{\ttfamily
  arXiv:1106.0701 [astro-ph.CO]}}.

\bibitem{Matias}
D.~Lopez~Nacir, R.~A. Porto, L.~Senatore, and M.~Zaldarriaga, ``{Dissipative
  effects in the Effective Field Theory of Inflation},''
  \href{http://dx.doi.org/10.1007/JHEP01(2012)075}{{\em JHEP} {\bfseries 01}
  (2012) 075}, \href{http://arxiv.org/abs/1109.4192}{{\ttfamily arXiv:1109.4192
  [hep-th]}}.

\bibitem{Bastero-Gil}
M.~Bastero-Gil, A.~Berera, I.~G. Moss, and R.~O. Ramos, ``{Theory of
  non-Gaussianity in warm inflation},''
  \href{http://dx.doi.org/10.1088/1475-7516/2014/12/008}{{\em JCAP} {\bfseries
  12} (2014) 008}, \href{http://arxiv.org/abs/1408.4391}{{\ttfamily
  arXiv:1408.4391 [astro-ph.CO]}}.

\bibitem{Berghaus}
K.~V. Berghaus, P.~W. Graham, and D.~E. Kaplan, ``{Minimal Warm Inflation},''
  \href{http://dx.doi.org/10.1088/1475-7516/2020/03/034}{{\em JCAP} {\bfseries
  03} (2020) 034}, \href{http://arxiv.org/abs/1910.07525}{{\ttfamily
  arXiv:1910.07525 [hep-ph]}}.

\bibitem{Yokoyama}
J.~Yokoyama and A.~D. Linde, ``{Is warm inflation possible?},''
  \href{http://dx.doi.org/10.1103/PhysRevD.60.083509}{{\em Phys. Rev. D}
  {\bfseries 60} (1999) 083509},
  \href{http://arxiv.org/abs/hep-ph/9809409}{{\ttfamily arXiv:hep-ph/9809409}}.

\bibitem{DeRocco}
W.~DeRocco, P.~W. Graham, and S.~Kalia, ``{Warming up cold inflation},''
  \href{http://dx.doi.org/10.1088/1475-7516/2021/11/011}{{\em JCAP} {\bfseries
  11} (2021) 011}, \href{http://arxiv.org/abs/2107.07517}{{\ttfamily
  arXiv:2107.07517 [hep-ph]}}.

\bibitem{Arnold_strike}
P.~B. Arnold and L.~D. McLerran, ``{The Sphaleron Strikes Back},''
  \href{http://dx.doi.org/10.1103/PhysRevD.37.1020}{{\em Phys. Rev. D}
  {\bfseries 37} (1988) 1020}.

\bibitem{Trodden}
M.~Trodden, ``{Electroweak baryogenesis},''
  \href{http://dx.doi.org/10.1103/RevModPhys.71.1463}{{\em Rev. Mod. Phys.}
  {\bfseries 71} (1999) 1463--1500},
  \href{http://arxiv.org/abs/hep-ph/9803479}{{\ttfamily arXiv:hep-ph/9803479}}.

\bibitem{Rubakov_sph}
V.~A. Rubakov and M.~E. Shaposhnikov, ``{Electroweak baryon number
  nonconservation in the early universe and in high-energy collisions},''
  \href{http://dx.doi.org/10.1070/PU1996v039n05ABEH000145}{{\em Usp. Fiz. Nauk}
  {\bfseries 166} (1996) 493--537},
  \href{http://arxiv.org/abs/hep-ph/9603208}{{\ttfamily arXiv:hep-ph/9603208}}.

\bibitem{Laine}
M.~Laine and A.~Vuorinen,
  \href{http://dx.doi.org/10.1007/978-3-319-31933-9}{{\em {Basics of Thermal
  Field Theory}}}, vol.~925.
\newblock Springer, 2016.
\newblock \href{http://arxiv.org/abs/1701.01554}{{\ttfamily arXiv:1701.01554
  [hep-ph]}}.

\bibitem{Arnold}
P.~B. Arnold, D.~Son, and L.~G. Yaffe, ``{The Hot baryon violation rate is O
  (alpha-w**5 T**4)},'' \href{http://dx.doi.org/10.1103/PhysRevD.55.6264}{{\em
  Phys. Rev. D} {\bfseries 55} (1997) 6264--6273},
  \href{http://arxiv.org/abs/hep-ph/9609481}{{\ttfamily arXiv:hep-ph/9609481}}.

\bibitem{Moore}
G.~D. Moore and M.~Tassler, ``{The Sphaleron Rate in SU(N) Gauge Theory},''
  \href{http://dx.doi.org/10.1007/JHEP02(2011)105}{{\em JHEP} {\bfseries 02}
  (2011) 105}, \href{http://arxiv.org/abs/1011.1167}{{\ttfamily arXiv:1011.1167
  [hep-ph]}}.

\bibitem{Moss_attractor}
I.~G. Moss and C.~Xiong, ``{On the consistency of warm inflation},''
  \href{http://dx.doi.org/10.1088/1475-7516/2008/11/023}{{\em JCAP} {\bfseries
  11} (2008) 023}, \href{http://arxiv.org/abs/0808.0261}{{\ttfamily
  arXiv:0808.0261 [astro-ph]}}.

\bibitem{Maldacena}
J.~M. Maldacena, ``{Non-Gaussian features of primordial fluctuations in single
  field inflationary models},''
  \href{http://dx.doi.org/10.1088/1126-6708/2003/05/013}{{\em JHEP} {\bfseries
  05} (2003) 013},
\href{http://arxiv.org/abs/astro-ph/0210603}{{\ttfamily arXiv:astro-ph/0210603
  [astro-ph]}}.
%%CITATION = ASTRO-PH/0210603;%%.

\bibitem{GW}
M.~Mirbabayi, L.~Senatore, E.~Silverstein, and M.~Zaldarriaga, ``{Gravitational
  Waves and the Scale of Inflation},''
  \href{http://dx.doi.org/10.1103/PhysRevD.91.063518}{{\em Phys. Rev. D}
  {\bfseries 91} (2015) 063518},
  \href{http://arxiv.org/abs/1412.0665}{{\ttfamily arXiv:1412.0665 [hep-th]}}.

\bibitem{Qiu}
Y.~Qiu and L.~Sorbo, ``{Spectrum of tensor perturbations in warm inflation},''
  \href{http://dx.doi.org/10.1103/PhysRevD.104.083542}{{\em Phys. Rev. D}
  {\bfseries 104} no.~8, (2021) 083542},
  \href{http://arxiv.org/abs/2107.09754}{{\ttfamily arXiv:2107.09754
  [astro-ph.CO]}}.

\bibitem{Klose}
P.~Klose, M.~Laine, and S.~Procacci, ``{Gravitational wave background from
  non-Abelian reheating after axion-like inflation},''
  \href{http://dx.doi.org/10.1088/1475-7516/2022/05/021}{{\em JCAP} {\bfseries
  05} (2022) 021}, \href{http://arxiv.org/abs/2201.02317}{{\ttfamily
  arXiv:2201.02317 [hep-ph]}}.

\bibitem{Planck}
{\bfseries Planck} Collaboration, Y.~Akrami {\em et~al.}, ``{Planck 2018
  results. IX. Constraints on primordial non-Gaussianity},''
  \href{http://dx.doi.org/10.1051/0004-6361/201935891}{{\em Astron. Astrophys.}
  {\bfseries 641} (2020) A9}, \href{http://arxiv.org/abs/1905.05697}{{\ttfamily
  arXiv:1905.05697 [astro-ph.CO]}}.

\bibitem{Babich}
D.~Babich, P.~Creminelli, and M.~Zaldarriaga, ``{The Shape of
  non-Gaussianities},''
  \href{http://dx.doi.org/10.1088/1475-7516/2004/08/009}{{\em JCAP} {\bfseries
  08} (2004) 009}, \href{http://arxiv.org/abs/astro-ph/0405356}{{\ttfamily
  arXiv:astro-ph/0405356}}.

\bibitem{Liu}
T.~Liu, X.~Tong, Y.~Wang, and Z.-Z. Xianyu, ``{Probing P and CP Violations on
  the Cosmological Collider},''
  \href{http://dx.doi.org/10.1007/JHEP04(2020)189}{{\em JHEP} {\bfseries 04}
  (2020) 189}, \href{http://arxiv.org/abs/1909.01819}{{\ttfamily
  arXiv:1909.01819 [hep-ph]}}.

\bibitem{Flauger}
R.~Flauger, M.~Mirbabayi, L.~Senatore, and E.~Silverstein, ``{Productive
  Interactions: heavy particles and non-Gaussianity},''
  \href{http://dx.doi.org/10.1088/1475-7516/2017/10/058}{{\em JCAP} {\bfseries
  10} (2017) 058}, \href{http://arxiv.org/abs/1606.00513}{{\ttfamily
  arXiv:1606.00513 [hep-th]}}.

\bibitem{Smith}
M.~M\"unchmeyer and K.~M. Smith, ``{Higher N-point function data analysis
  techniques for heavy particle production and WMAP results},''
  \href{http://dx.doi.org/10.1103/PhysRevD.100.123511}{{\em Phys. Rev. D}
  {\bfseries 100} no.~12, (2019) 123511},
  \href{http://arxiv.org/abs/1910.00596}{{\ttfamily arXiv:1910.00596
  [astro-ph.CO]}}.

\bibitem{Creminelli_eq}
P.~Creminelli, A.~Nicolis, L.~Senatore, M.~Tegmark, and M.~Zaldarriaga,
  ``{Limits on non-gaussianities from wmap data},''
  \href{http://dx.doi.org/10.1088/1475-7516/2006/05/004}{{\em JCAP} {\bfseries
  05} (2006) 004}, \href{http://arxiv.org/abs/astro-ph/0509029}{{\ttfamily
  arXiv:astro-ph/0509029}}.

\bibitem{Shahin}
A.~Maleknejad, M.~M. Sheikh-Jabbari, and J.~Soda, ``{Gauge Fields and
  Inflation},'' \href{http://dx.doi.org/10.1016/j.physrep.2013.03.003}{{\em
  Phys. Rept.} {\bfseries 528} (2013) 161--261},
  \href{http://arxiv.org/abs/1212.2921}{{\ttfamily arXiv:1212.2921 [hep-th]}}.

\bibitem{Adshead}
P.~Adshead and M.~Wyman, ``{Chromo-Natural Inflation: Natural inflation on a
  steep potential with classical non-Abelian gauge fields},''
  \href{http://dx.doi.org/10.1103/PhysRevLett.108.261302}{{\em Phys. Rev.
  Lett.} {\bfseries 108} (2012) 261302},
  \href{http://arxiv.org/abs/1202.2366}{{\ttfamily arXiv:1202.2366 [hep-th]}}.

\bibitem{Kamali}
V.~Kamali, ``{Warm pseudoscalar inflation},''
  \href{http://dx.doi.org/10.1103/PhysRevD.100.043520}{{\em Phys. Rev. D}
  {\bfseries 100} no.~4, (2019) 043520},
  \href{http://arxiv.org/abs/1901.01897}{{\ttfamily arXiv:1901.01897 [gr-qc]}}.

\bibitem{Anber}
M.~M. Anber and L.~Sorbo, ``{Naturally inflating on steep potentials through
  electromagnetic dissipation},''
  \href{http://dx.doi.org/10.1103/PhysRevD.81.043534}{{\em Phys. Rev. D}
  {\bfseries 81} (2010) 043534},
  \href{http://arxiv.org/abs/0908.4089}{{\ttfamily arXiv:0908.4089 [hep-th]}}.

\bibitem{trapped}
D.~Green, B.~Horn, L.~Senatore, and E.~Silverstein, ``{Trapped Inflation},''
  \href{http://dx.doi.org/10.1103/PhysRevD.80.063533}{{\em Phys. Rev. D}
  {\bfseries 80} (2009) 063533},
  \href{http://arxiv.org/abs/0902.1006}{{\ttfamily arXiv:0902.1006 [hep-th]}}.

\bibitem{Abolhasani}
A.~A. Abolhasani and M.~M. Sheikh-Jabbari, ``{Resonant reconciliation of
  convex-potential inflation models and the Planck data},''
  \href{http://dx.doi.org/10.1103/PhysRevD.100.103505}{{\em Phys. Rev. D}
  {\bfseries 100} no.~10, (2019) 103505},
  \href{http://arxiv.org/abs/1903.05120}{{\ttfamily arXiv:1903.05120
  [astro-ph.CO]}}.

\bibitem{Ferreira}
R.~Z. Ferreira and A.~Notari, ``{Thermalized Axion Inflation},''
  \href{http://dx.doi.org/10.1088/1475-7516/2017/09/007}{{\em JCAP} {\bfseries
  09} (2017) 007}, \href{http://arxiv.org/abs/1706.00373}{{\ttfamily
  arXiv:1706.00373 [astro-ph.CO]}}.

\bibitem{Domcke}
V.~Domcke, Y.~Ema, and K.~Mukaida, ``{Chiral Anomaly, Schwinger Effect,
  Euler-Heisenberg Lagrangian, and application to axion inflation},''
  \href{http://dx.doi.org/10.1007/JHEP02(2020)055}{{\em JHEP} {\bfseries 02}
  (2020) 055}, \href{http://arxiv.org/abs/1910.01205}{{\ttfamily
  arXiv:1910.01205 [hep-ph]}}.

\bibitem{Rubakov}
D.~Y. Grigoriev, V.~A. Rubakov, and M.~E. Shaposhnikov, ``{TOPOLOGICAL
  TRANSITIONS AT FINITE TEMPERATURES: A REAL TIME NUMERICAL APPROACH},''
  \href{http://dx.doi.org/10.1016/0550-3213(89)90553-1}{{\em Nucl. Phys. B}
  {\bfseries 326} (1989) 737--757}.

\end{thebibliography}\endgroup
\end{document}